\documentclass[pre,twocolumn,aps]{revtex4}
\newcommand{\br}{{\bf r}}
\newcommand{\bR}{{\bf R}}
\newcommand{\rR}{\mathfrak{R}}

\newcommand{\bF}{{\bf F}}
\newcommand{\rF}{{\rm F}}

\newcommand{\tbR}{\tilde{\bf R}}

\usepackage[english]{babel}
\usepackage{graphicx}
\usepackage{comment}
\usepackage{bm}
\usepackage{amsmath,amssymb} 
\usepackage{epstopdf}
\usepackage{verbatim}
\usepackage{float}

\begin{document}

\title{$NVU$ perspective on simple liquids' quasiuniversality}
\author{Jeppe C. Dyre}
\email{dyre@ruc.dk}
\affiliation{DNRF Centre ``Glass and Time'', IMFUFA, Department of Sciences, Roskilde University, Postbox 260, DK-4000 Roskilde, Denmark}

\date{\today}

\begin{abstract}
The last half century of research into the structure, dynamics, and thermodynamics of simple liquids has revealed a number of approximate universalities. This paper argues that simple liquids' reduced-coordinate constant-potential-energy hypersurfaces constitute a quasiuniversal family of compact Riemannian manifolds parameterized by a single number, from which follows these liquids' quasiuniversalities.
\end{abstract}
\maketitle

\section{Introduction}\label{sec1}
Simple liquids are traditionally defined by contrast to complex liquids as systems of spherically symmetric particles interacting via pair forces 
\cite{pit55,ber64,fis64,ric65,tem68,ros79a,ail80,rowlinson,gubbins,chandler,barrat,deb05,han06,dou07,kir07,bag10}. 
It is now known, however, that a number of such systems like the Gaussian core model \cite{sti76,sti97,pre05,zac08}, the Lennard-Jones Gaussian model \cite{eng07}, the Jagla and related discontinuous-force models \cite{jag99,yan06,lom07,fra07}, and other models \cite{deo06,fra07,bar09,bar11,for11} exhibit quite complex behavior. On the other hand, van der Waals {\it molecular} liquids are generally regular and ``simple'' in their properties \cite{barrat,chandler}. In view of these facts we recently with Ingebrigtsen and Schr{\o}der suggested \cite{ing12} defining instead liquid simplicity from the property of strong correlations between equilibrium virial and potential energy fluctuations in the $NVT$ ensemble \cite{I,II,III,IV,V}. This is how the term ``simple liquid'' is used below. In practice there is considerable overlap between the two definitions, for instance the Lennard-Jones liquid and related systems are simple in both senses. One notable difference, however, is that realistic liquids are only simple in the present meaning of the term in part of their phase diagram -- simplicity does not apply near or at the critical point or at gas states (where different kinds of simplicity apply, of course). With regard to real liquids, it appears that most or all van der Waals bonded and metallic liquids are simple, whereas covalently bonded, hydrogen-bonded, and strongly ionic liquids are generally not simple because competing interactions weaken the virial potential-energy correlations \cite{ing12}.

Simple liquids are characterized by having isomorphs in their phase diagram \cite{IV}. An isomorph is an equivalence class of the following equivalence relation: two state points are isomorphic if all pairs of physically relevant microconfigurations of the state points, which trivially scale into one another, have the same configuration-space canonical probability. Only inverse-power-law (IPL) potentials have exact isomorphs, but all ``strongly correlating'' liquids have isomorphs to a good approximation \cite{IV}. 

Simple liquids' simple properties derive from the fact that the existence of isomorphs implies that their thermodynamic phase diagram is effectively one-dimensional instead of two-dimensional for the several properties that are isomorph invariant. Examples of such properties are \cite{IV}: Newtonian and Brownian reduced-unit dynamics, reduced-unit static structure factors of any order, the excess entropy, the isochoric heat capacity. For any simple liquid melting defines an isomorph in the phase diagram; this implies invariance along the melting curve of, e.g., excess entropy, reduced viscosity, reduced heat conductivity, reduced diffusion constant, etc, as well as invariance of the Lindemann melting criterion \cite{IV,V}. 

The isomorph theory explains a number of previously noted regularities relating to a given simple liquid \cite{IV}. The theory cannot explain, however, the intriguing similarities between {\it different} simple liquids known for a long time. This is the focus of the present paper that views simple-liquid quasiuniversality from an $NVU$ perspective. 

$NVU$ dynamics \cite{NVU_I}, which is inspired by earlier approaches to dynamics conserving \cite{cot86,sca02,cot06} or limiting \cite{wan07} the potential energy, is defined as geodesic motion on the constant-potential-energy hypersurface. If $\bR\equiv(\br_1,...,\br_N)$ is the $3N$-dimensional position vector describing a system of $N$ particles, $\bR_i$ the position vector at time step $i$, and $\bF_i$ the corresponding $3N$-dimensional force vector, the $NVU$ algorithm \cite{NVU_I} is $\bR_{i+1}=2\bR_i-\bR_{i-1}-2 [\bF_i\cdot\left( \bR_i-\bR_{i-1}\right)]\bF_i/\bF_i^2$. As shown in Ref. \onlinecite{NVU_II}, if $m$ is the particle mass and $\Delta t$ the time step of the Verlet algorithm $\bR_{i+1}=2\bR_i-\bR_{i-1}+ \bF_i(\Delta t)^2/m$, the $NVU$ and Verlet algorithms are equivalent in the thermodynamic limit because the fluctuations of the $NVU$-force prefactor become insignificant as $N\rightarrow\infty$. Consequently, the radial distribution function, diffusion constant, coherent and incoherent intermediate scattering functions, etc, are identical in the thermodynamic limits of $NVU$ and standard $NVE$ or $NVT$ Newtonian dynamics. This has been confirmed in computer simulations of both atomic and molecular models \cite{NVU_I,NVU_II,NVU_III}.

$NVU$ dynamics provides an alternative view of a liquid's molecular dynamics. At any given state point all information about the liquid's structure and dynamics is encoded in its constant-potential-energy hypersurface $\Omega$. If $\langle U\rangle$ is the average potential energy, this compact Riemannian differentiable manifold is defined by

\begin{equation}\label{om}
\Omega 
\,\equiv\,\{\bR\,|\,\, U(\bR)=\langle U\rangle\}\,.
\end{equation}
$\Omega$ is a so-called level set of the function $U(\bR)$. If standard periodic boundary conditions are employed, $\Omega$ is embedded as a $(3N-1)$-dimensional hypersurface in the $3N$-dimensional torus. The manifold $\Omega$ is not only defined for simple liquids, of course, but for all liquids, as well as for all solids and gasses. 

The present paper considers systems of $N$ identical particles in volume $V$. For simplicity we focus on systems interacting via pairwise additive forces. Thermodynamic quantities are excess quantities, i.e., in excess of the corresponding ideal gas quantities at the same density and temperature. Thus $S$ is the {\it excess} extensive entropy ($S<0$) and $C_V$ the {\it excess} extensive isochoric specific heat, which we for simplicity refer to as just ``entropy'' and ``specific heat''. The corresponding intensive quantities are denoted by lower-case letters, i.e., $s\equiv S/N$ and $c_V\equiv C_V/N$. Reduced quantities are marked by a tilde.

It is clear from the results of many years of research into the liquid state that there is no such thing as exact universality, even among narrowly restricted classes like the IPL liquids. Thus any theory predicting liquid-state universality is too simple. The philosophy of this paper is that approximate universalities may provide useful insights. In this connection it is an obvious conjecture that genuine simple-liquid universality is approached as the spatial dimension increases towards infinity -- if this is the case, we have for liquids a situation reminiscent to that of critical phenomena.

Section \ref{sec2} argues for the existence of a quasiuniversal family of constant-potential-energy hypersurfaces for simple liquids, parameterized by just one parameter. The argument presented is not rigorous, but suggests one route for justifying quasiuniversality. The reader may choose to accept quasiuniversality in its $NVU$ formulation, skip most of Sec. \ref{sec2}, and proceed to the central part of the paper, Sec. \ref{sec3}, which derives and discusses a number of consequences of the $NVU$ formulation of simple liquids' quasiuniversality. Section \ref{sec4} returns briefly to the question what causes quasiuniversality. Finally, Sec. \ref{sec5} gives a few concluding remarks.

\section{A single-parameter family of hypersurfaces $\tilde\Omega(\lambda)$ common to all simple liquids} \label{sec2}

Any state point in the thermodynamic phase diagram of a liquid gives rise to a constant-potential-energy hypersurface $\Omega$ as defined in Eq. (\ref{om}). At first sight these manifolds may appear to be completely characterized by the number $\langle U\rangle$, but actually the system volume $V$ is a second parameter implicit in the definition of $U(\bR)$. Thus for liquids in general, $\Omega$ is described by two parameters, corresponding to two independent thermodynamic variables. 

Using reduced units means measuring length in units of $\rho^{-1/3}$, time in units of $\rho^{-1/3}\sqrt{m/k_BT}$, and energy in units of $k_BT$. The reduced $3N$-dimensional position vector is thus defined by $\tbR\equiv\rho^{1/3}\bR$. Appendix A of  Ref.\onlinecite{IV} showed that a liquid has strong virial potential-energy correlations (i.e., is simple) if and only if the liquid has isomorphs (to a good approximation), and that this happens if and only if the liquid's reduced-unit constant-potential-energy hypersurfaces are (almost) invariant along certain curves in the phase diagram (the isomorphs). Thus for any given simple liquid a single number, $\lambda$, parameterizes the reduced-coordinate constant-potential-energy hypersurfaces. We indicate this by writing

\begin{equation}\label{omt0}
\tilde\Omega
\,=\,\tilde\Omega(\lambda)
\end{equation}
where

\begin{equation}\label{omt}
\tilde\Omega
\,\equiv\,\{\tbR\,|\,\, U(\rho^{-1/3}\tbR)=\langle U\rangle\}\,.
\end{equation}
The isomorph theory says nothing about how these hypersurfaces compare between different simple liquids. We now argue that the family $\tilde\Omega(\lambda)$ is quasiuniversal, i.e., approximately the same for all simple liquids. 

The only systems with 100\% correlation between $NVT$ equilibrium fluctuations of potential energy and virial ($W(\bR)\equiv -1/3\sum_i\br_i\cdot\nabla_i U$) are the inverse-power-law (IPL) systems, for which the potential energy scales with interparticle distance as $\propto r^{-n}$. For these systems the 100\% correlation follows from the identity $U(\lambda\bR)=\lambda^{-n}U(\bR)$ and Euler's theorem for homogeneous functions. The last decade has given rise to a number of studies of how well experimental systems may be understood by reference to an IPL system (see, e.g., Refs. \onlinecite{nga05,cos08,cos09a,rol10,gammagamma}). In these works the value of the IPL exponent $n$ generally plays a central role because $n$ determines how the system reacts to volume changes.

A simple liquid by definition has strong $WU$ correlations at its condensed-phase state points. Consequently, at each of it's strongly correlating state points the liquid behaves much like an IPL system. The value of the effective IPL exponent $n$ generally varies with state point. For Lennard-Jones (LJ) liquids at typical state points $n$ is fairly constant, between 5 and 6. For the 99.9\% correlating ``repulsive LJ liquid'' defined by the pair potential $\epsilon[(r/\sigma)^{-12}+(r/\sigma)^{-6}]/2$ \cite{ing12a}, the exponent varies from $n\cong 12$ at high densities to $n\cong 6$ at low densities. How much the effective IPL exponent varies throughout the phase diagram is of no importance for the arguments given below for $\tilde\Omega(\lambda)$ quasiuniversality, however.

In many respects IPL systems with different exponents have similar (``quasiuniversal'') behavior. This was noted long time ago in relation to these systems' structure and DC dynamic properties expressed, e.g., via the  diffusion constant \cite{ash66,han69,gro71,hoo71,hiw74,sti75,ros76,kan85,ind94}. During the last decade IPL quasiuniversality has come into focus again \cite{bra06,hey07,hey08} and been extended to include general dynamic properties, termed ``dynamic equivalence'' by Medina-Noyola and coworkers. Dynamic equivalence has been established for Brownian \cite{gue03,lop11a} as well as Newtonian \cite{you03,you05,pon11a,sch11} dynamics. 

Because structure and dynamics are both encoded in $\tilde\Omega$, IPL quasiuniversality follows if IPL systems have almost identical constant-potential-energy hypersurfaces. But why should this be the case? To understand this, we consider two infinitesimally close configurations with same $n$-IPL potential energy and show that they for all $m$ to a good approximation have the same $m$-IPL potential energy. 

The $m$-IPL potential energy is given by $U_m=\varepsilon_m\sum_{ij} (r_{ij}/\sigma)^{-m}$, in which $r_{ij}$ is the distance between particles $i$ and $j$. The change in $U_m$ between two nearby configurations is given by 

\begin{equation}\label{ddn0}
\delta U_m
\,=\,-m\,\varepsilon_m\sum_{ij}  \left(\frac{r_{ij}}{\sigma}\right)^{-m-1}\frac{\delta r_{ij}}{\sigma}\,.
\end{equation}
By assumption $\delta U_n=0$. Equation (\ref{ddn0}) implies

\begin{equation}\label{ddn}
\frac{d}{dm}\left(\frac{\delta U_m}{m\,\varepsilon_m}\right)
\,=\,\sum_{ij}  \left(\frac{r_{ij}}{\sigma}\right)^{-m-1}\ln\left(\frac{r_{ij}}{\sigma}\right)\frac{\delta r_{ij}}{\sigma}\,.
\end{equation}
The factor $\ln(r_{ij}/\sigma)$ in Eq. (\ref{ddn}) does not vary much because it is a logarithm. This factor is multiplied by $(r_{ij}/\sigma)^{-m-1}$ that varies a lot. Their product is dominated by a rather narrow range of interparticle distances, the most important of which is denoted by $\langle r\rangle_m$. This quantity depends on both $m$ and the state point. As a good approximation one can replace $\ln(r_{ij}/\sigma)$ by $\ln (\langle r\rangle_m/\sigma)$, and Eq. (\ref{ddn}) now becomes via Eq. (\ref{ddn0})

\begin{equation}\label{ddn2}
\frac{d}{d m}\left(\frac{\delta U_m}{m\,\varepsilon_m}\right)
\,=\,-\ln \left(\frac{\langle r\rangle_m}{\sigma}\right)\left(\frac{\delta U_m}{m\,\varepsilon_m}\right)\,.
\end{equation}
Recall that $\delta U_n=0$. Since $\delta U_m /(m\varepsilon_m)\equiv 0$ is the unique solution to this first-order differential equation that obeys $\delta U_n/(n\varepsilon_n)=0$, it follows that $\delta U_m=0$ for all $m$. -- Note that the approximation made by replacing $\ln(r_{ij}/\sigma)$ by a constant is questionable when $m\leq 2$, in which case no narrow range of interparticle distances dominates Eq. (\ref{ddn}) because $\sum_jr_{ij}^{-m-1}$ diverges. Note also that, since the approximation relies on fluctuations in $r_{ij}$ being small in the region where force is greatest, the argument can be expected to work better in higher dimensions. This supports the conjecture that genuine simple-liquid universality is approached as the spatial dimension increases towards infinity (Sec. \ref{sec1}). 

Within the above approximation the potential-energy functions $U_n(\bR)$ have the same constant-potential-energy hypersurfaces for all $n\ge 2$. Possibly, this applies for all $n$. Since a simple liquid at a given state point may be approximated by an IPL system, this means that all simple liquids have approximately the same constant-potential-energy hypersurfaces. In other words, to a good approximation a family of manifolds parameterized by a single parameter, $\tilde\Omega(\lambda)$, is common to all simple liquids. Each isomorph of a simple liquid corresponds to a particular value of $\lambda$, i.e., to one specific manifold $\tilde\Omega$.

\section{Consequences of $\tilde\Omega(\lambda)$ quasiuniversality}\label{sec3}

This section derives consequences of the above justified basic idea that the family of reduced-coordinate constant-potential-energy hypersurfaces to a good approximation is common to all simple liquids.

\subsection{Different IPL systems exhibit close similarities with respect to structure and dynamics, similarities that extend to all other simple liquids \cite{ash66,han69,gro71,hoo71,hiw74,sti75,ros76,kan85,ind94,nor00,you03,you05,gue03,spe03,she04,hey05,sco05,bra06,hey07,hey08,ore08,hey09,lan09,ram09,sch11,pon11a,lop11a,ram11}}

Because reduced-unit structure and dynamics are both encoded in $\tilde\Omega$, any two state points of two different IPL systems with same $\tilde\Omega$ have the same structure and dynamics. The extension of similarities to all simple liquids follows from $\tilde\Omega(\lambda)$ quasiuniversality. Note that the hard-sphere (HS) system also exhibits these quasiuniversalities because it is the $n\rightarrow\infty$ limit of $n$-IPL systems.

\subsection{The Young-Andersen approximate scaling principle \cite{you03,you05}}

This principle states that if two liquids at two state points have the same reduced-unit radial distribution function $g(\tilde r)$, they have the same reduced-unit dynamics. $g(\tilde r)$ is determined from $\tilde\Omega$, so having the same $g(\tilde r)$ implies having the same $\tilde\Omega$. This implies the same dynamics.

\subsection{Quasiuniversality of the order-parameter maps of Debenedetti and coworkers \cite{tru00,err03,cha07}}

Plotting a translational order parameter versus an orientational order parameter for various state points leads to a one-dimensional curve for any simple liquid, because both order parameters are isomorph invariant \cite{IV}. The approximate identity between the order-parameter curves of different simple liquids follows from $\tilde\Omega(\lambda)$ quasiuniversality, because $\tilde\Omega$ determines both order parameters.

\subsection{Excess entropy scaling \cite{ros77,ros99,pon11,sin12}}

Rosenfeld noted in 1977 that the reduced-unit diffusion constants $\tilde D$ of different simple liquids have an approximately universal dependence on the entropy per particle, $s$ \cite{ros77}. This quasiuniversality applies also, e.g., for the heat conductivity as a function of excess entropy \cite{gro85}. For any simple liquid, since $\tilde D$ and $s$ are both isomorph invariant, one quantity is a function of the other. Quasiuniversality of the function $\tilde D(s)$ is a consequence of the fact that $\tilde D$ and $s$ are both encoded in $\tilde\Omega$ (the entropy $S=Ns$ is the logarithm of the area of $\tilde\Omega$).

\subsection{The Lindemann melting criterion \cite{gil56,ubb65,ros69}}

According to the Lindemann criterion a crystal melts when the vibrational mean-square displacement obeys $\sqrt{\langle\tilde x^2\rangle}\simeq 0.1$, where $\tilde x\equiv x\rho^{1/3}$ in which $x$ is the atomic vibrational displacement from equilibrium in an axis direction. The melting curve in the phase diagram is an isomorph \cite{IV,V}, so melting takes place for a particular manifold $\tilde\Omega_c$ of the crystalline state. This manifold determines $\langle\tilde x^2\rangle$. Thus any simple crystal melts when $\langle\tilde x^2\rangle$ reaches a certain, quasiuniversal value. The Lindemann criterion and its generalizations \cite{gil56,ros69,sai06} have been questioned on the grounds that they are single-phase criteria, whereas melting occurs when the crystal and liquid free energies are the same, so any melting criterion should refer to properties of both phases. One possible resolution of this paradox is that the Lindemann criterion does not, in fact, determine the melting line, but a spinodal at a slightly higher temperature where the crystal becomes mechanically unstable \cite{mal00,st}. Alternatively, for the class of simple liquids $\tilde\Omega$ quasiuniversality implies that there is basically just one melting process, which takes place at the state point where the crystalline manifold is $\tilde\Omega_c$. Any single-phase melting criterion referring to this manifold applies for all simple liquids.

\subsection{Freezing rules referring to the liquid} 

Quasiuniversality of such rules follow from the fact that $\tilde\Omega$ is quasiuniversal also on the liquid side of melting. For instance, this implies the Hansen-Verlet rule that a liquid crystallizes when the first peak of the radial distribution function reaches the value 2.85 \cite{han69}. Likewise, any simple liquid's $c_V$ is close to $3k_B$ at freezing \cite{wallace,bol12} (it is shown below that $c_V$ is encoded in $\tilde\Omega$). Other quasiuniversal melting rules similarly follow from $\tilde\Omega(\lambda)$ quasiuniversality \cite{sai06}. Examples are the Andrade equation from 1934 predicting a quasiuniversal value of the reduced-unit melting point viscosity \cite{and34,kap05}, the Raveche-Mountain-Streett criterion \cite{rav74} of a quasiuniversal ratio between maximum and minimum of the radial distribution function at freezing, Lyapunov-exponent based criteria \cite{mal00}, or the criterion of zero higher-than-second-order liquid configurational entropy at crystallization \cite{sai01}. Note that the theory also predicts a quasiuniversal constant-volume melting entropy for simple liquids, which is consistent with experiment \cite{tal80,wallace}.

\subsection{Algebraic closedness of the class of simple potentials}\label{sec3g}

If $U_1(\bR)$ and $U_2(\bR)$ are both potentials of simple liquids, i.e., strongly correlating, their sum and product are also strongly correlating potentials: since $U_1(\tilde\bR)$ and $U_2(\tilde\bR)$ are both constant on the manifolds $\tilde\Omega(\lambda)$, this applies also for their sum and product. In particular, note the following property. Writing a simple pair potentials as $v(r)=\varepsilon\phi(r/\sigma)$, the derivative with respect to $\sigma$, $\partial v(r)/\partial\sigma$, is also a simple pair potential. Less trivial is the property that the product of two pair potentials of simple liquids defines the pair potential of a simple liquid (see Sec. \ref{sec4}).

\subsection{Additivity of thermodynamic quantities}

Suppose $U(\bR)=U_1(\bR)\pm U_2(\bR)$ in which $U_1(\bR)$ and $U_2(\bR)$ each define a simple liquid (here and below the symbol $\pm$ signals that the arguments apply for both signs). An example is when $U(\bR)$ is the LJ potential, $U_1(\bR)$ is an $n=12$ IPL pair potential and $U_2(\bR)$ is an $n=6$ IPL pair potential. Then $U(\bR)$ defines a simple liquid, and as functions of density and entropy the corresponding temperatures obey $T(\rho,S)=T_1(\rho,S)\pm T_2(\rho,S)$ \cite{ros82}. To show this, note that since the entropy determines $\tilde\Omega$, at given values of $\rho$ and $S$ the three constant-potential-energy manifolds are identical: $\Omega=\Omega_1=\Omega_2$. This implies that $U(\rho,S)=U_1(\rho,S)\pm U_2(\rho,S)$, from which $T(\rho,S)=T_1(\rho,S)\pm T_2(\rho,S)$ follows via the definition of temperature $T\equiv(\partial U/\partial S)_\rho$. The thermodynamic relation $W=(\partial U/\partial \ln\rho)_S$ similarly implies additivity of virials: $W(\rho,S)=W_1(\rho,S)\pm W_2(\rho,S)$. The (excess) Helmholtz free energy $F$, (excess) Gibbs free energy $G$, and (excess) enthalpy $H$ are likewise additive: $F(\rho,S)=F_1(\rho,S)\pm F_2(\rho,S)$, $G(\rho,S)=G_1(\rho,S)\pm G_2(\rho,S)$, $H(\rho,S)=H_1(\rho,S)\pm H_2(\rho,S)$.

As an application we note the intriguing ``additivity of melting temperatures'' first discussed by Rosenfeld \cite{ros76}: Since crystallization for all simple liquids takes place at a certain value of the liquid entropy, at any given density one has $T_m=T_{m,1}\pm T_{m,2}$. An IPL liquid's melting temperature scales with density as $T_m\propto \rho^{n/3}$, so for the LJ liquid this implies an expression of the form $T_m = A \rho^4- B\rho^2$ \cite{ros76,khr11,ing12a}.

\subsection{A partly quasiuniversal equation of state} 

It was recently shown that simple liquids have simple thermodynamics in the sense that temperature factorizes into a product of a function of entropy and a function of density, $T=f(s)h(\rho)$ \cite{ing12a}. We now show that the function $f(s)$ is quasiuniversal, i.e., all specific system dependence is in the function $h(\rho)$. This justifies writing the equation of state as

\begin{equation}\label{eos}
T\,=\,f_0(s)h(\rho)\,.
\end{equation}
The point is that the specific heat, like the entropy, depends only on $\tilde\Omega$. This can be shown by first writing $c_V$ in terms of fluctuations of canonical ensemble probabilities, and then relating the latter to microcanonical (NVU) probabilities, arguing as follows. According to Einstein $C_V=\langle (\Delta U)^2\rangle/k_ BT^2$ in which the average refers to the canonical ensemble. In terms of the configuration-space probability $p\propto\exp(-U/k_BT)$ this implies $C_V=k_B\langle(\Delta\ln p)^2\rangle$. The canonical ensemble is realized from the microcanonical $NVU$ ensemble in the standard textbook way by considering a small subvolume $V_m$ of the total volume $V$. On average $V_m$ contains $m$ particles where $m/N=V_m/V$. Each configuration of $m$ particles in $V_m$, $(\br_1,...,\br_m)$, has a probability $p(\br_1,...,\br_m)$ that can be calculated from the manifold $\Omega$ (or $\tilde\Omega$)  by integrating out the remaining degrees of freedom. The set of configurations in $\Omega$ with precisely $m$ particles in volume $V_m$ is denoted by $\Omega_m$. Integrating out the remaining degrees of freedom from the configurations in $\Omega_m$ determines $p(\br_1,...,\br_m)$, so this function is given by $\Omega$ (or $\tilde\Omega$). From $c_V=k_B\langle(\Delta\ln p)^2\rangle/m$ it now follows that $c_V=c_V(\lambda)$. Since $c_V=(\partial s/\partial\ln T)_\rho$ and $s=s(\lambda)$, this implies that at fixed density $d\ln T=\phi_0(\lambda)d\lambda$ for some quasiuniversal function $\phi_0(\lambda)$. Thus, while for two simple liquids the temperatures corresponding to the same manifold $\tilde\Omega$ may well differ, the relative temperature changes (at fixed density) between different $\tilde\Omega$s are the same. By integration this implies that for each simple liquid one can write $T=\Phi(\lambda)T_\ast(\rho)$. Combining this with the equation of state $T=f(s)h(\rho)$ shows that the function $f(s)$ is determined by $\lambda$, i.e., by the manifold $\tilde\Omega$. In summary, $\tilde\Omega(\lambda)$ quasiuniversality implies that $f(s)$ is quasiuniversal, $f(s)=f_0(s)$. The function $T_\ast(\rho)=h(\rho)$ is not quasiuniversal; it reflects how the liquid's characteristic energy scale varies with density \cite{alb02}.

\subsection{Quasiuniversality of simple liquids' specific-heat temperature dependence}

Eliminating $\lambda$ between $c_V(\lambda)$ and $T=f_0(s)h(\rho)$ where $s=s(\lambda)$ leads to $c_V=F_0(T/h(\rho))$ for a quasiuniversal function $F_0$. This is consistent with the Rosenfeld-Tarazona expression $c_V\propto T^{-2/5}$ \cite{ros99}, which as shown by computer simulations applies to a good approximation not only for all IPL systems, but also for LJ-type liquids and other simple liquids \cite{sci99,dol03,yan04,geb05,ped10,trond}. Note that $c_V\propto T^{-2/5}$ implies $s\propto -T^{-2/5}$ since $c_V=(\partial s/\partial\ln T)_\rho$ and $s\rightarrow 0$ for $T\rightarrow\infty$. This means that 

\begin{equation}\label{f0}
f_0(s)
\,\propto\, (-s)^{-5/2}\,.
\end{equation}

\subsection{Quasiuniversal isochoric fragility of simple liquids}

$NVU$ dynamics give the same relaxation times as $NVE$ or $NVT$ dynamics, and thus the reduced-unit relaxation time $\tilde\tau$ is determined by $\tilde\Omega$. This not only means that $\tilde\tau$ is a unique function of the excess entropy (``excess entropy scaling''), it also implies a quasiuniversal temperature dependence of $\tilde\tau$ at constant density: The quasiuniversal equation of state Eq. (\ref{eos}) implies that at any given density, entropy is a quasiuniversal function of temperature in the following sense: $s=s_0(T/h(\rho))$. This implies quasiuniversality of the form $\tilde\tau=\tilde\tau_0(T/h(\rho))$ \cite{alb02}. In particular, at constant density Angell's fragility, $-d\log_{10}(\tilde\tau)/d\ln T|_{T=T_g}$ \cite{ang95}, is a quasiuniversal number for any given cooling rate defining the glass transition temperature $T_g$ \cite{joh74,ang95,dyr06}. This is reminiscent of the universal temperature dependence of viscosity discussed in 1996 by Kivelson {\it et al.} \cite{kiv96}, although these authors subtracted the high-temperature activation energy before demonstrating data collapse. A quasiuniversal isochoric fragility is consistent with simulations of De Michele {\it et al.}, who found that different IPL systems have the same fragility \cite{dem04} (see, however, also Ref. \onlinecite{sen11}). The prediction is not entirely consistent with available experimental data, although there does seem to be a tendency that van der Waals liquids have isochoric fragilities not far from 50 \cite{nis07}. In this connection it should be pointed out that it is an experimental challenge to determine the isochoric fragility accurately.

\subsection{The hard-sphere system}

Temperature plays no role for the configurational degrees of freedom of the hard-sphere (HS) liquid. Since only density is important, the HS thermodynamic phase diagram is effectively one-dimensional. This brings to mind isomorphs, the existence of which implies that a simple liquid's phase diagram is also effectively one-dimensional. Is the HS liquid simple? Since its potential energy is zero whereas the virial is not, the HS liquid is not simple in the sense of the term used here. In our opinion, the HS liquid should be thought of more as the $n\rightarrow\infty$ limit of an $n$-IPL system than as a physical system of its own right. When a simple liquid is modeled by a HS system, each of the liquid's isomorphs corresponds to a specific value of the HS packing fraction $\eta$. This establishes a one-to-one correspondence $\lambda\leftrightarrow\eta$, which explains why simple liquids' entropy, relaxation time, viscosity, etc, have all been found to be quasiuniversal functions of the $\eta$ parameter of the HS reference system.

Arguments for quasiuniversality were traditionally based on the fact that any simple liquid is well represented by the HS reference system \cite{sco05,ber64,zwa54,lon64,wid67,wca,tor10}. In this view, the HS system is useful because it captures the essence of liquids' harsh repulsive forces \cite{wid67,wca,bri09}. This picture is intuitively appealing, but runs into problems when confronted with known facts. On the one hand, IPL quasiuniversality extends down to $n=3$ or $n=4$, in fact for some quantities  down to $n=1$ \cite{ros77,ros98a} where repulsions are quite smooth. On the other hand, there are several systems with harsh repulsive forces that exhibit anomalous behavior which is not captured by the HS system \cite{eng07,jag99,yan06,lom07,deo06,fra07,bar09,bar11,for11}. From the $NVU$ perspective, the HS system's usefulness is not the {\it explanation} of simple liquids' quasiuniversality, but a {\it consequence} of it: since all $n$-IPL systems are quasiuniversal, the HS system inherits this property because it is the $n\rightarrow\infty$ limit of $n$-IPL systems.

\subsection{Role of entropy}

Theories relating entropy to a liquid's relaxation time go back in time at least to Bestul and Chang, who in 1964 noted that the glass transitions of different glass-forming liquids occur at virtually the same value of the (excess) entropy \cite{bes64}. Since the glass transition for a given cooling rate takes place when the liquid's relaxation time reaches a certain value, by generalization to other cooling rates this result implies that the relaxation time is a quasiuniversal function of entropy. Independently, based on computer simulations and analytical arguments, Rosenfeld in 1977 proposed excess entropy scaling \cite{ros77}. These two results, as well as the Adam-Gibbs model from 1965 in which entropy is also  crucial \cite{ada65,dud08,dyr09}, may appear counterintuitive since entropy is global property: How can a global property control the relaxation time, which is determined as an average of local properties? For simple liquids $\tilde\Omega(\lambda)$ quasiuniversality provides the following answer. Entropy {\it identifies} the relevant manifold $\tilde\Omega$,  and $\tilde\Omega$ {\it determines} the relaxation time. Accordingly, other unique markers of $\tilde\Omega$ should be equally useful for determining the relaxation time, for instance the two-particle entropy that Dzugutov in 1996 suggested controls the relaxation time \cite{sai06,dzu96}.

\subsection{Characterizing $\tilde\Omega$ via the mean curvature}

It is difficult to visualize a high-dimensional differentiable manifold. A primitive analog is a two-dimensional closed surface in ordinary three-dimensional space. Such a surface has two obvious characteristics, its area and its mean curvature. The latter is conveniently quantified in terms of the average radius of curvature $\rR$, defined as the average inverse curvature. Both the area and the curvature concepts generalize to multidimensional Riemannian surfaces \cite{hicks,dom68,riemannian} (good introductions to this branch of mathematics are available, for instance Refs. \onlinecite{panoramic,gol05,surface}). The entropy is the logarithm of the manifold's area, but what is the physical interpretation of the average radius of curvature? To answer this we start from the configuration-space canonical ensemble expression \cite{LLstat,rug97,pow05}, 

\begin{equation}\label{ct}
k_BT
\,=\,\frac{\langle (\nabla U)^2\rangle}{\langle \nabla^2 U\rangle}\,,
\end{equation}
which is derived by partial integration of $\int d\bR \nabla\cdot\nabla U(\bR) \exp[-U(\bR)/k_BT]$. Because of ensemble equivalence, the $NVU$ configuration-space microcanonical ensemble may be used to calculate the averages in Eq. (\ref{ct}) as integrals over $\Omega$. The inverse radius of curvature at a point on a $d$-dimensional hypersurface $\Omega$ is $\nabla\cdot {\bf n}/d$ \cite{gol05,surface} where $\bf n$ is the normal vector to $\Omega$ at the point that is in our case given by $\bf n = \nabla U/|\nabla U|$. To leading order in $1/\sqrt{N}$ fluctuations are small and variations in the denominator are insignificant. This implies for the inverse average radius of curvature $1/\rR$ (replacing $3N-1$ by $3N$)

\begin{equation}\label{r}
\frac{1}{\rR}
\,=\,\frac{1}{3N}\frac{\langle \nabla^2 U\rangle}{\langle | \nabla U |\rangle}\,.
\end{equation}
If the average length of the $3N$-dimensional force vector $\bF=-\nabla U$ is denoted by $\rF$, because fluctuations are insignificant as $N\rightarrow\infty$, one has $\rF^2=\langle(\nabla U)^2\rangle$ and Eqs. (\ref{ct}) and (\ref{r}) imply

\begin{equation}\label{trf1}
\rR\,\rF
\,=\,3Nk_BT
\,.
\end{equation}
We see that a small radius of curvature of $\Omega$ corresponds physically to a large average force. As $N\rightarrow\infty$, $\langle \nabla^2 U\rangle\sim N$ and $\langle | \nabla U |\rangle\sim \sqrt N$ which implies $\rR\sim \sqrt N$. Likewise, $\rF\sim \sqrt N$ as $N\rightarrow\infty$. If one defines $\tilde \rR\equiv \rho^{1/3}\rR/\sqrt{3N}$ and $\tilde \rF\equiv \rho^{-1/3}\rF/(\sqrt{3N}k_BT)$, these quantities are dimensionless, independent of $N$ in the thermodynamic limit, and related by

\begin{equation}\label{trf2}
\tilde\rR\,\tilde\rF
\,=\,1\,.
\end{equation}
Thus $\tilde\Omega$'s curvature is basically $\tilde\rF$. This quantity provides an alternative to the entropy for characterizing $\tilde\Omega$. 

Entropy and $\tilde\rF$ both have simple geometric interpretations, but the curvature $\tilde\rF$ has the advantage of being the average of a locally defined quantity. This implies that, since fluctuations are unimportant in the thermodynamic limit, $\tilde\rF$ may be calculated from a short-time simulation. Note also that $\tilde\rF$ may be calculated from standard $NVE$ or $NVT$ simulations. 

The quasiuniversal entropy dependencies observed for simple liquids' structure and dynamics may equally well be interpreted as quasiuniversal dependencies on $\tilde\rF$, the curvature of $\tilde\Omega$. Interestingly, this is the quantity that controls the relaxation time in the entropic barrier hopping theory of Schweizer and co-workers \cite{sch07,tri09,tri11} -- except for the fact that here the real forces are replaced by effective forces defined by the direct correlation function.

The connection between curvature and force is not surprising since motion on a flat manifold requires no force. Indeed, this point was made by one of the pioneers in connecting mechanics and differential geometry, Lipschitz, who wrote in 1873 \cite{lip1873}: ``When a material particle, which is not influenced by any accelerating force, is bound to move on a given surface, the pressure exerted in each point of the trajectory is inversely proportional to the radius of curvature of this trajectory'' (quoted from Ref. \onlinecite{lut95}). What happens is the following. When a particle (i.e., the system) moves on a perfectly smooth surface like $\tilde\Omega$, since no work is performed, the kinetic energy is conserved and thus the particle's velocity $v$ is constant. This implies that the centripetal force keeping the particle on the surface, $\propto v^2/r$, is inversely proportional to the local curvature radius $r$.

For viscous liquids most motion is vibrational and one can estimate $\tilde\rF$ by adopting a harmonic approximation. Writing for the force on a particle $-Cx$, where $x$ is the displacement away from its short-time average, implies  $\langle\rF^2\rangle \propto C^2\langle x^2\rangle$. Since $C\langle x^2\rangle/2=k_BT/2$ by equipartition, this means that if $a\equiv\rho^{-1/3}$ is the average interatomic spacing, $\langle\tilde\rF^2 \rangle\equiv \rho^{-2/3} \langle\rF^2\rangle/[3N(k_BT)^2]\propto a^2/\langle x^2\rangle\equiv 1/\langle \tilde x^2\rangle$. Thus for viscous simple liquids the quantity $a^2/\langle x^2\rangle$ identifies $\tilde\Omega$ and, in this sense, ``controls'' the relaxation time in the same sense as entropy does. This is an old idea \cite{fly68,hal87,koh88,buc92,heu94,sok94,san98,sta02}, which is closely related to the reasoning behind the shoving model and other elastic models for the temperature dependence of viscous liquids' relaxation time \cite{dyr96,dyr06,lar08,dyr12}. Note also that, since $1/\langle \tilde x^2\rangle$ identifies $\tilde\Omega$, for simple liquids the crossover to activated transitions takes place at a quasiuniversal value of the reduced-unit vibrational displacement $\langle \tilde x^2\rangle$, as recently predicted by Lubchenko and coworkers from the random first-order transition theory \cite{lub07,rab12}.

\subsection{Quasiuniversal interdependence of isomorph invariants}

We showed above that the isomorph invariant $c_V$ is encoded in $\tilde\Omega$. Generally, any isomorph invariant is encoded in $\tilde\Omega$. As a consequence, $\tilde\Omega(\lambda)$ quasiuniversality implies the following principle:

\begin{itemize}
\item {The relation between any two isomorph invariants is quasiuniversal.}
\end{itemize}
Thus from the $\tilde\Omega(\lambda)$ quasiuniversality perspective, excess entropy scaling is a special case of a much more general principle.

\subsection{A single microconfiguration is enough to identify $\tilde\Omega$}

Given an equilibrium configuration $\bR=(\br_1,...,\br_N)$ the corresponding reduced-coordinate vector $\tilde\bR$ identifies the relevant manifold $\tilde\Omega$. From $\tilde\Omega$ all the system's isomorph invariants like entropy, specific heat, reduced relaxation time, reduced diffusion constant, reduced incoherent scattering function, etc, can be calculated. The reduced-unit radial distribution function $g(\tilde r)$ is also included, of course, being trivially given by $\tilde\bR$. This illustrates again the Young-Andersen approximate scaling principle that for simple liquids knowing $g(\tilde r)$ determines many other quantities.

\section{Towards a theory of simple liquids}\label{sec4}

Rosenfeld in 1977 justified IPL quasiuniversality by arguing that each n-IPL system is well represented by a HS reference system \cite{ros77}. Since the HS system has just one parameter, this implies IPL quasiuniversality. Given the fact that the HS system is the $n\rightarrow\infty$ limit of n-IPL systems, this reasoning may be regarded as circular by assuming part of what is to be arrived at. Section \ref{sec2} gave one argument for quasiuniversality, but we wish here to supplement it by another argument, suggesting a deeper reason for quasiuniversality.

We take as starting point that the pair potential defined by a simple exponentially decaying function of $r$ is strongly correlating, i.e., defines a simple liquid. This remains to be thoroughly investigated and documented, but the recent simulations by Veldhorst {\it et al.} \cite{vel12} of the Buckingham potential, which has a harsh exponentially repulsive term, certainly indicates that this is the case. Thus we assume that systems with pair potentials of the form $v(r)=\varepsilon\exp(-r/\sigma)$ are strongly correlating, at least in a significant part of the $(\sigma^3\rho,k_BT/\varepsilon)$ parameter space. This means that the reduced-coordinate constant-potential-energy hypersurfaces $\tilde\Omega$, which are {\it a priori} parameterized by the two dimensionless numbers $\sigma^3\rho$ and $k_BT/\varepsilon$, constitute a single-parameter family $\tilde\Omega(\lambda)$, where $\lambda=\lambda(\sigma^3\rho,k_BT/\varepsilon)$. 

We proceed to argue that this is the quasiuniversal one-parameter family characterizing simple liquids. It is enough to show that all IPL pair potentials have in common these ``exponential'' reduced-coordinate constant-potential-energy hypersurfaces. Now, following the reasoning of Sec. \ref{sec3g} any linear combination of exponential pair potentials, $C_1\exp(-r/\sigma_1)+C_2\exp(-r/\sigma_2)$, has the same family of reduced-coordinate constant-potential-energy hypersurfaces as a single exponential. By generalization, this implies via the mathematical identity

\begin{equation}\label{IPL}
r^{-n}
\,=\,\frac{1}{(n-1)!}\int_0^\infty x^{n-1}e^{-xr}dx
\end{equation}
that the family $\tilde\Omega(\lambda)$ is common to the IPL potentials and, by implication, to all simple liquids. Thus, while the IPL functions constitute a convenient ``basis set'' for simple-liquid pair potentials, an even simpler basis set is provided by the set of exponential pair potentials. Note that such potentials come out naturally from quantum mechanics; in fact, these were the first pair potentials discussed in the literature \cite{smi97} (by Born, Born-Meyer, Morse, etc).

A consequence of the above is also that the product $v(r)\propto v_1(r)v_2(r)$ of two simple-liquid pair potentials defines a simple liquid. This is because each of the functions $v_1(r)$ and $v_2(r)$ are sums of exponentials, and consequently so is $v(r)$.

\section{Concluding remarks}\label{sec5}

We have shown that the several quasiuniversalities found for simple liquids' structure, dynamics, and thermodynamics follow from one fundamental quasiuniversality, namely the existence of a common single-parameter family of reduced-coordinate constant-potential-energy hypersurfaces $\tilde\Omega(\lambda)$. This ``$\tilde\Omega(\lambda)$ quasiuniversality'' was justified for IPL systems in Sec. \ref{sec2} by a non-rigorous argument, but it can also be arrived at assuming that the exponentially repulsive pair potential is simple (Sec. \ref{sec4}). From this quasiuniversality generalizes to all simple liquids by virtue of their property of having strong correlations between $NVT$ equilibrium virial and potential-energy fluctuations.

It is important to emphasize again that there is no exact universality among simple liquids, only {\it approximate} universality. A clear demonstration of this is provided by the well-known fact that the crystalline state is only face-centered cubic for IPL exponents larger than seven (below which it is body-centered cubic). Another point to be emphasized is that, in contrast to mode-coupling theory and other so-called fully renormalized theories of liquid dynamics, the present approach does not distinguish between short and long time scales -- quasiuniversality applies on vibrational time scales and longer.

\acknowledgments 

The author is indebted to Nick Bailey, Jean-Louis Barrat, Charu Chakravarty, Jack Douglas, Holger Bech Nielsen, Thomas Schr{\o}der, Ken Schweizer, Tom Truskett, and Ben Widom, for providing constructive critiques of early versions of the manuscript, as well as inspiration and encouragement. The centre for viscous liquid dynamics ``Glass and Time'' is sponsored by the Danish National Research Foundation (DNRF).

\bibliography{jcd}

\begin{thebibliography}{154}
\expandafter\ifx\csname natexlab\endcsname\relax\def\natexlab#1{#1}\fi
\expandafter\ifx\csname bibnamefont\endcsname\relax
  \def\bibnamefont#1{#1}\fi
\expandafter\ifx\csname bibfnamefont\endcsname\relax
  \def\bibfnamefont#1{#1}\fi
\expandafter\ifx\csname citenamefont\endcsname\relax
  \def\citenamefont#1{#1}\fi
\expandafter\ifx\csname url\endcsname\relax
  \def\url#1{\texttt{#1}}\fi
\expandafter\ifx\csname urlprefix\endcsname\relax\def\urlprefix{URL }\fi
\providecommand{\bibinfo}[2]{#2}
\providecommand{\eprint}[2][]{\url{#2}}

\bibitem[{\citenamefont{Pitzer}(1955)}]{pit55}
\bibinfo{author}{\bibfnamefont{K.~S.} \bibnamefont{Pitzer}},
  \bibinfo{journal}{J. Am. Chem. Soc.} \textbf{\bibinfo{volume}{77}},
  \bibinfo{pages}{3427} (\bibinfo{year}{1955}).

\bibitem[{\citenamefont{Bernal}(1964)}]{ber64}
\bibinfo{author}{\bibfnamefont{J.~D.} \bibnamefont{Bernal}},
  \bibinfo{journal}{Proc. R. Soc. London Ser. A}
  \textbf{\bibinfo{volume}{280}}, \bibinfo{pages}{299} (\bibinfo{year}{1964}).

\bibitem[{\citenamefont{Fisher}(1964)}]{fis64}
\bibinfo{author}{\bibfnamefont{I.~Z.} \bibnamefont{Fisher}},
  \emph{\bibinfo{title}{Statistical Theory of Liquids}}
  (\bibinfo{publisher}{University of Chicago, Chicago}, \bibinfo{year}{1964}).

\bibitem[{\citenamefont{Rice and Gray}(1965)}]{ric65}
\bibinfo{author}{\bibfnamefont{S.~A.} \bibnamefont{Rice}} \bibnamefont{and}
  \bibinfo{author}{\bibfnamefont{P.}~\bibnamefont{Gray}},
  \emph{\bibinfo{title}{The Statistical Mechanics of Simple Liquids}}
  (\bibinfo{publisher}{Interscience, New York}, \bibinfo{year}{1965}).

\bibitem[{\citenamefont{Temperley et~al.}(1968)\citenamefont{Temperley,
  Rowlinson, and Rushbrooke}}]{tem68}
\bibinfo{author}{\bibfnamefont{H.~N.~V.} \bibnamefont{Temperley}},
  \bibinfo{author}{\bibfnamefont{J.~S.} \bibnamefont{Rowlinson}},
  \bibnamefont{and} \bibinfo{author}{\bibfnamefont{G.~S.}
  \bibnamefont{Rushbrooke}}, \emph{\bibinfo{title}{Physics of Simple Liquids}}
  (\bibinfo{publisher}{Wiley}, \bibinfo{year}{1968}).

\bibitem[{\citenamefont{Rosenfeld and Ashcroft}(1979)}]{ros79a}
\bibinfo{author}{\bibfnamefont{Y.}~\bibnamefont{Rosenfeld}} \bibnamefont{and}
  \bibinfo{author}{\bibfnamefont{N.~W.} \bibnamefont{Ashcroft}},
  \bibinfo{journal}{Phys. Lett. A} \textbf{\bibinfo{volume}{73}},
  \bibinfo{pages}{31} (\bibinfo{year}{1979}).

\bibitem[{\citenamefont{Ailawadi}(1980)}]{ail80}
\bibinfo{author}{\bibfnamefont{N.~K.} \bibnamefont{Ailawadi}},
  \bibinfo{journal}{Phys. Rep.} \textbf{\bibinfo{volume}{57}},
  \bibinfo{pages}{241} (\bibinfo{year}{1980}).

\bibitem[{\citenamefont{Rowlinson and Widom}(1982)}]{rowlinson}
\bibinfo{author}{\bibfnamefont{J.~S.} \bibnamefont{Rowlinson}}
  \bibnamefont{and} \bibinfo{author}{\bibfnamefont{B.}~\bibnamefont{Widom}},
  \emph{\bibinfo{title}{Molecular Theory of Capillarity}}
  (\bibinfo{publisher}{Clarendon, Oxford}, \bibinfo{year}{1982}).

\bibitem[{\citenamefont{Gray and Gubbins}(1984)}]{gubbins}
\bibinfo{author}{\bibfnamefont{C.~G.} \bibnamefont{Gray}} \bibnamefont{and}
  \bibinfo{author}{\bibfnamefont{K.~E.} \bibnamefont{Gubbins}},
  \emph{\bibinfo{title}{Theory of Molecular Fluids}}
  (\bibinfo{publisher}{Oxford University Press}, \bibinfo{year}{1984}).

\bibitem[{\citenamefont{Chandler}(1987)}]{chandler}
\bibinfo{author}{\bibfnamefont{D.}~\bibnamefont{Chandler}},
  \emph{\bibinfo{title}{Introduction to Modern Statistical Mechanics}}
  (\bibinfo{publisher}{Oxford University Press}, \bibinfo{year}{1987}).

\bibitem[{\citenamefont{Barrat and Hansen}(2003)}]{barrat}
\bibinfo{author}{\bibfnamefont{J.~L.} \bibnamefont{Barrat}} \bibnamefont{and}
  \bibinfo{author}{\bibfnamefont{J.~P.} \bibnamefont{Hansen}},
  \emph{\bibinfo{title}{Basic Concepts for Simple and Complex Liquids}}
  (\bibinfo{publisher}{Cambridge University Press}, \bibinfo{year}{2003}).

\bibitem[{\citenamefont{Debenedetti}(2005)}]{deb05}
\bibinfo{author}{\bibfnamefont{P.~G.} \bibnamefont{Debenedetti}},
  \bibinfo{journal}{AICHE J.} \textbf{\bibinfo{volume}{51}},
  \bibinfo{pages}{2391} (\bibinfo{year}{2005}).

\bibitem[{\citenamefont{Hansen and McDonald}(2006)}]{han06}
\bibinfo{author}{\bibfnamefont{J.-P.} \bibnamefont{Hansen}} \bibnamefont{and}
  \bibinfo{author}{\bibfnamefont{I.~R.} \bibnamefont{McDonald}},
  \emph{\bibinfo{title}{Theory of Simple Liquids}}
  (\bibinfo{publisher}{Academic, New York}, \bibinfo{year}{2006}),
  \bibinfo{edition}{3rd} ed.

\bibitem[{\citenamefont{Douglas et~al.}({2007})\citenamefont{Douglas, Dudowicz,
  and Freed}}]{dou07}
\bibinfo{author}{\bibfnamefont{J.~F.} \bibnamefont{Douglas}},
  \bibinfo{author}{\bibfnamefont{J.}~\bibnamefont{Dudowicz}}, \bibnamefont{and}
  \bibinfo{author}{\bibfnamefont{K.~F.} \bibnamefont{Freed}},
  \bibinfo{journal}{J. Chem. Phys.} \textbf{\bibinfo{volume}{{127}}},
  \bibinfo{pages}{224901} (\bibinfo{year}{{2007}}).

\bibitem[{\citenamefont{Kirchner}(2007)}]{kir07}
\bibinfo{author}{\bibfnamefont{B.}~\bibnamefont{Kirchner}},
  \bibinfo{journal}{Phys. Rep.} \textbf{\bibinfo{volume}{440}},
  \bibinfo{pages}{1} (\bibinfo{year}{2007}).

\bibitem[{\citenamefont{Bagchi and Chakravarty}(2010)}]{bag10}
\bibinfo{author}{\bibfnamefont{B.}~\bibnamefont{Bagchi}} \bibnamefont{and}
  \bibinfo{author}{\bibfnamefont{C.}~\bibnamefont{Chakravarty}},
  \bibinfo{journal}{J. Chem. Sci.} \textbf{\bibinfo{volume}{122}},
  \bibinfo{pages}{459} (\bibinfo{year}{2010}).

\bibitem[{\citenamefont{Stillinger}(1976)}]{sti76}
\bibinfo{author}{\bibfnamefont{F.~H.} \bibnamefont{Stillinger}},
  \bibinfo{journal}{J. Chem. Phys.} \textbf{\bibinfo{volume}{65}},
  \bibinfo{pages}{3968} (\bibinfo{year}{1976}).

\bibitem[{\citenamefont{Stillinger and Stillinger}(1997)}]{sti97}
\bibinfo{author}{\bibfnamefont{F.~H.} \bibnamefont{Stillinger}}
  \bibnamefont{and} \bibinfo{author}{\bibfnamefont{D.~K.}
  \bibnamefont{Stillinger}}, \bibinfo{journal}{Physica A}
  \textbf{\bibinfo{volume}{244}}, \bibinfo{pages}{358} (\bibinfo{year}{1997}).

\bibitem[{\citenamefont{Prestipino et~al.}(2005)\citenamefont{Prestipino, Saja,
  and Giaquinta}}]{pre05}
\bibinfo{author}{\bibfnamefont{S.}~\bibnamefont{Prestipino}},
  \bibinfo{author}{\bibfnamefont{F.}~\bibnamefont{Saja}}, \bibnamefont{and}
  \bibinfo{author}{\bibfnamefont{P.~V.} \bibnamefont{Giaquinta}},
  \bibinfo{journal}{J. Chem. Phys.} \textbf{\bibinfo{volume}{123}},
  \bibinfo{pages}{144110} (\bibinfo{year}{2005}).

\bibitem[{\citenamefont{Zachary et~al.}(2008)\citenamefont{Zachary, Stillinger,
  and Torquato}}]{zac08}
\bibinfo{author}{\bibfnamefont{C.~E.} \bibnamefont{Zachary}},
  \bibinfo{author}{\bibfnamefont{F.~H.} \bibnamefont{Stillinger}},
  \bibnamefont{and} \bibinfo{author}{\bibfnamefont{S.}~\bibnamefont{Torquato}},
  \bibinfo{journal}{J. Chem. Phys.} \textbf{\bibinfo{volume}{128}},
  \bibinfo{pages}{224505} (\bibinfo{year}{2008}).

\bibitem[{\citenamefont{Engel and Trebin}(2007)}]{eng07}
\bibinfo{author}{\bibfnamefont{M.}~\bibnamefont{Engel}} \bibnamefont{and}
  \bibinfo{author}{\bibfnamefont{H.~R.} \bibnamefont{Trebin}},
  \bibinfo{journal}{Phys. Rev. Lett.} \textbf{\bibinfo{volume}{98}},
  \bibinfo{pages}{225505} (\bibinfo{year}{2007}).

\bibitem[{\citenamefont{Jagla}(1999)}]{jag99}
\bibinfo{author}{\bibfnamefont{E.~A.} \bibnamefont{Jagla}},
  \bibinfo{journal}{J. Chem. Phys.} \textbf{\bibinfo{volume}{111}},
  \bibinfo{pages}{8980} (\bibinfo{year}{1999}).

\bibitem[{\citenamefont{Yan et~al.}(2006)\citenamefont{Yan, Buldyrev,
  Giovambattista, Debenedetti, and Stanley}}]{yan06}
\bibinfo{author}{\bibfnamefont{Z.}~\bibnamefont{Yan}},
  \bibinfo{author}{\bibfnamefont{S.~V.} \bibnamefont{Buldyrev}},
  \bibinfo{author}{\bibfnamefont{N.}~\bibnamefont{Giovambattista}},
  \bibinfo{author}{\bibfnamefont{P.~G.} \bibnamefont{Debenedetti}},
  \bibnamefont{and} \bibinfo{author}{\bibfnamefont{H.~E.}
  \bibnamefont{Stanley}}, \bibinfo{journal}{Phys. Rev. E}
  \textbf{\bibinfo{volume}{73}}, \bibinfo{pages}{051204}
  (\bibinfo{year}{2006}).

\bibitem[{\citenamefont{Lomba et~al.}(2007)\citenamefont{Lomba, Almarza,
  Martin, and McBride}}]{lom07}
\bibinfo{author}{\bibfnamefont{E.}~\bibnamefont{Lomba}},
  \bibinfo{author}{\bibfnamefont{N.~G.} \bibnamefont{Almarza}},
  \bibinfo{author}{\bibfnamefont{C.}~\bibnamefont{Martin}}, \bibnamefont{and}
  \bibinfo{author}{\bibfnamefont{C.}~\bibnamefont{McBride}},
  \bibinfo{journal}{J. Chem. Phys.} \textbf{\bibinfo{volume}{126}},
  \bibinfo{pages}{244510 } (\bibinfo{year}{2007}).

\bibitem[{\citenamefont{Franzese}(2007)}]{fra07}
\bibinfo{author}{\bibfnamefont{G.}~\bibnamefont{Franzese}},
  \bibinfo{journal}{J. Mol. Liq.} \textbf{\bibinfo{volume}{136}},
  \bibinfo{pages}{267} (\bibinfo{year}{2007}).

\bibitem[{\citenamefont{de~Oliveira et~al.}(2006)\citenamefont{de~Oliveira,
  Netz, Colla, and Barbosa}}]{deo06}
\bibinfo{author}{\bibfnamefont{A.~B.} \bibnamefont{de~Oliveira}},
  \bibinfo{author}{\bibfnamefont{P.~A.} \bibnamefont{Netz}},
  \bibinfo{author}{\bibfnamefont{T.}~\bibnamefont{Colla}}, \bibnamefont{and}
  \bibinfo{author}{\bibfnamefont{M.~C.} \bibnamefont{Barbosa}},
  \bibinfo{journal}{J. Chem. Phys.} \textbf{\bibinfo{volume}{125}},
  \bibinfo{pages}{124503} (\bibinfo{year}{2006}).

\bibitem[{\citenamefont{Barraz et~al.}(2009)\citenamefont{Barraz, Salcedo, and
  Barbosa}}]{bar09}
\bibinfo{author}{\bibfnamefont{N.~M.} \bibnamefont{Barraz}},
  \bibinfo{author}{\bibfnamefont{E.}~\bibnamefont{Salcedo}}, \bibnamefont{and}
  \bibinfo{author}{\bibfnamefont{M.~C.} \bibnamefont{Barbosa}},
  \bibinfo{journal}{J. Chem. Phys.} \textbf{\bibinfo{volume}{131}},
  \bibinfo{pages}{094504} (\bibinfo{year}{2009}).

\bibitem[{\citenamefont{Barraz et~al.}(2011)\citenamefont{Barraz, Salcedo, and
  Barbosa}}]{bar11}
\bibinfo{author}{\bibfnamefont{N.~M.} \bibnamefont{Barraz}},
  \bibinfo{author}{\bibfnamefont{E.}~\bibnamefont{Salcedo}}, \bibnamefont{and}
  \bibinfo{author}{\bibfnamefont{M.~C.} \bibnamefont{Barbosa}},
  \bibinfo{journal}{J. Chem. Phys.} \textbf{\bibinfo{volume}{135}},
  \bibinfo{pages}{104507} (\bibinfo{year}{2011}).

\bibitem[{\citenamefont{Fomin et~al.}(2011)\citenamefont{Fomin, Tsiok, and
  Ryzhov}}]{for11}
\bibinfo{author}{\bibfnamefont{Y.~D.} \bibnamefont{Fomin}},
  \bibinfo{author}{\bibfnamefont{E.~N.} \bibnamefont{Tsiok}}, \bibnamefont{and}
  \bibinfo{author}{\bibfnamefont{V.~N.} \bibnamefont{Ryzhov}},
  \bibinfo{journal}{J. Chem. Phys.} \textbf{\bibinfo{volume}{135}},
  \bibinfo{pages}{234502} (\bibinfo{year}{2011}).

\bibitem[{\citenamefont{Ingebrigtsen
  et~al.}(2012{\natexlab{a}})\citenamefont{Ingebrigtsen, Schr\o{}der, and
  Dyre}}]{ing12}
\bibinfo{author}{\bibfnamefont{T.~S.} \bibnamefont{Ingebrigtsen}},
  \bibinfo{author}{\bibfnamefont{T.~B.} \bibnamefont{Schr\o{}der}},
  \bibnamefont{and} \bibinfo{author}{\bibfnamefont{J.~C.} \bibnamefont{Dyre}},
  \bibinfo{journal}{Phys. Rev. X} \textbf{\bibinfo{volume}{2}},
  \bibinfo{pages}{011011} (\bibinfo{year}{2012}{\natexlab{a}}).

\bibitem[{\citenamefont{Bailey et~al.}(2008{\natexlab{a}})\citenamefont{Bailey,
  Pedersen, Gnan, Schr{\o}der, and Dyre}}]{I}
\bibinfo{author}{\bibfnamefont{N.~P.} \bibnamefont{Bailey}},
  \bibinfo{author}{\bibfnamefont{U.~R.} \bibnamefont{Pedersen}},
  \bibinfo{author}{\bibfnamefont{N.}~\bibnamefont{Gnan}},
  \bibinfo{author}{\bibfnamefont{T.~B.} \bibnamefont{Schr{\o}der}},
  \bibnamefont{and} \bibinfo{author}{\bibfnamefont{J.~C.} \bibnamefont{Dyre}},
  \bibinfo{journal}{J. Chem. Phys.} \textbf{\bibinfo{volume}{129}},
  \bibinfo{pages}{184507} (\bibinfo{year}{2008}{\natexlab{a}}).

\bibitem[{\citenamefont{Bailey et~al.}(2008{\natexlab{b}})\citenamefont{Bailey,
  Pedersen, Gnan, Schr{\o}der, and Dyre}}]{II}
\bibinfo{author}{\bibfnamefont{N.~P.} \bibnamefont{Bailey}},
  \bibinfo{author}{\bibfnamefont{U.~R.} \bibnamefont{Pedersen}},
  \bibinfo{author}{\bibfnamefont{N.}~\bibnamefont{Gnan}},
  \bibinfo{author}{\bibfnamefont{T.~B.} \bibnamefont{Schr{\o}der}},
  \bibnamefont{and} \bibinfo{author}{\bibfnamefont{J.~C.} \bibnamefont{Dyre}},
  \bibinfo{journal}{J. Chem. Phys.} \textbf{\bibinfo{volume}{129}},
  \bibinfo{pages}{184508} (\bibinfo{year}{2008}{\natexlab{b}}).

\bibitem[{\citenamefont{Schr{\o}der et~al.}(2009)\citenamefont{Schr{\o}der,
  Bailey, Pedersen, Gnan, and Dyre}}]{III}
\bibinfo{author}{\bibfnamefont{T.~B.} \bibnamefont{Schr{\o}der}},
  \bibinfo{author}{\bibfnamefont{N.~P.} \bibnamefont{Bailey}},
  \bibinfo{author}{\bibfnamefont{U.~R.} \bibnamefont{Pedersen}},
  \bibinfo{author}{\bibfnamefont{N.}~\bibnamefont{Gnan}}, \bibnamefont{and}
  \bibinfo{author}{\bibfnamefont{J.~C.} \bibnamefont{Dyre}},
  \bibinfo{journal}{J. Chem. Phys.} \textbf{\bibinfo{volume}{131}},
  \bibinfo{pages}{234503} (\bibinfo{year}{2009}).

\bibitem[{\citenamefont{Gnan et~al.}(2009)\citenamefont{Gnan, Schr{\o}der,
  Pedersen, Bailey, and Dyre}}]{IV}
\bibinfo{author}{\bibfnamefont{N.}~\bibnamefont{Gnan}},
  \bibinfo{author}{\bibfnamefont{T.~B.} \bibnamefont{Schr{\o}der}},
  \bibinfo{author}{\bibfnamefont{U.~R.} \bibnamefont{Pedersen}},
  \bibinfo{author}{\bibfnamefont{N.~P.} \bibnamefont{Bailey}},
  \bibnamefont{and} \bibinfo{author}{\bibfnamefont{J.~C.} \bibnamefont{Dyre}},
  \bibinfo{journal}{J. Chem. Phys.} \textbf{\bibinfo{volume}{131}},
  \bibinfo{pages}{234504} (\bibinfo{year}{2009}).

\bibitem[{\citenamefont{Schr{\o}der et~al.}(2011)\citenamefont{Schr{\o}der,
  Gnan, Pedersen, Bailey, and Dyre}}]{V}
\bibinfo{author}{\bibfnamefont{T.~B.} \bibnamefont{Schr{\o}der}},
  \bibinfo{author}{\bibfnamefont{N.}~\bibnamefont{Gnan}},
  \bibinfo{author}{\bibfnamefont{U.~R.} \bibnamefont{Pedersen}},
  \bibinfo{author}{\bibfnamefont{N.~P.} \bibnamefont{Bailey}},
  \bibnamefont{and} \bibinfo{author}{\bibfnamefont{J.~C.} \bibnamefont{Dyre}},
  \bibinfo{journal}{J. Chem. Phys.} \textbf{\bibinfo{volume}{134}},
  \bibinfo{pages}{164505} (\bibinfo{year}{2011}).

\bibitem[{\citenamefont{Ingebrigtsen
  et~al.}(2011{\natexlab{a}})\citenamefont{Ingebrigtsen, Toxvaerd, Heilmann,
  Schr{\o}der, and Dyre}}]{NVU_I}
\bibinfo{author}{\bibfnamefont{T.~S.} \bibnamefont{Ingebrigtsen}},
  \bibinfo{author}{\bibfnamefont{S.}~\bibnamefont{Toxvaerd}},
  \bibinfo{author}{\bibfnamefont{O.~J.} \bibnamefont{Heilmann}},
  \bibinfo{author}{\bibfnamefont{T.~B.} \bibnamefont{Schr{\o}der}},
  \bibnamefont{and} \bibinfo{author}{\bibfnamefont{J.~C.} \bibnamefont{Dyre}},
  \bibinfo{journal}{J. Chem. Phys.} \textbf{\bibinfo{volume}{135}},
  \bibinfo{pages}{104101} (\bibinfo{year}{2011}{\natexlab{a}}).

\bibitem[{\citenamefont{Cotterill and Madsen}(1986)}]{cot86}
\bibinfo{author}{\bibfnamefont{R.~M.~J.} \bibnamefont{Cotterill}}
  \bibnamefont{and} \bibinfo{author}{\bibfnamefont{J.~U.}
  \bibnamefont{Madsen}}, \bibinfo{journal}{Phys. Rev. B}
  \textbf{\bibinfo{volume}{33}}, \bibinfo{pages}{262} (\bibinfo{year}{1986}).

\bibitem[{\citenamefont{Scala et~al.}(2002)\citenamefont{Scala, Angelani, {Di
  Leonardo}, Ruocco, and Sciortino}}]{sca02}
\bibinfo{author}{\bibfnamefont{A.}~\bibnamefont{Scala}},
  \bibinfo{author}{\bibfnamefont{L.}~\bibnamefont{Angelani}},
  \bibinfo{author}{\bibfnamefont{R.}~\bibnamefont{{Di Leonardo}}},
  \bibinfo{author}{\bibfnamefont{G.}~\bibnamefont{Ruocco}}, \bibnamefont{and}
  \bibinfo{author}{\bibfnamefont{F.}~\bibnamefont{Sciortino}},
  \bibinfo{journal}{Phil. Mag. B} \textbf{\bibinfo{volume}{82}},
  \bibinfo{pages}{151} (\bibinfo{year}{2002}).

\bibitem[{\citenamefont{Cotterill and Madsen}(2006)}]{cot06}
\bibinfo{author}{\bibfnamefont{R.~M.~J.} \bibnamefont{Cotterill}}
  \bibnamefont{and} \bibinfo{author}{\bibfnamefont{J.~U.}
  \bibnamefont{Madsen}}, \bibinfo{journal}{J. Phys.: Condens. Matter}
  \textbf{\bibinfo{volume}{18}}, \bibinfo{pages}{6507} (\bibinfo{year}{2006}).

\bibitem[{\citenamefont{Wang and Stratt}(2007)}]{wan07}
\bibinfo{author}{\bibfnamefont{C.}~\bibnamefont{Wang}} \bibnamefont{and}
  \bibinfo{author}{\bibfnamefont{R.~M.} \bibnamefont{Stratt}},
  \bibinfo{journal}{J. Chem. Phys.} \textbf{\bibinfo{volume}{127}},
  \bibinfo{pages}{224503} (\bibinfo{year}{2007}).

\bibitem[{\citenamefont{Ingebrigtsen
  et~al.}(2011{\natexlab{b}})\citenamefont{Ingebrigtsen, Toxvaerd, Schr{\o}der,
  and Dyre}}]{NVU_II}
\bibinfo{author}{\bibfnamefont{T.~S.} \bibnamefont{Ingebrigtsen}},
  \bibinfo{author}{\bibfnamefont{S.}~\bibnamefont{Toxvaerd}},
  \bibinfo{author}{\bibfnamefont{T.~B.} \bibnamefont{Schr{\o}der}},
  \bibnamefont{and} \bibinfo{author}{\bibfnamefont{J.~C.} \bibnamefont{Dyre}},
  \bibinfo{journal}{J. Chem. Phys.} \textbf{\bibinfo{volume}{135}},
  \bibinfo{pages}{104102} (\bibinfo{year}{2011}{\natexlab{b}}).

\bibitem[{\citenamefont{Ingebrigtsen and Dyre}(2012)}]{NVU_III}
\bibinfo{author}{\bibfnamefont{T.~S.} \bibnamefont{Ingebrigtsen}}
  \bibnamefont{and} \bibinfo{author}{\bibfnamefont{J.~C.} \bibnamefont{Dyre}},
  \bibinfo{journal}{J. Chem. Phys.} \textbf{\bibinfo{volume}{137}},
  \bibinfo{pages}{244101} (\bibinfo{year}{2012}).

\bibitem[{\citenamefont{Ngai et~al.}(2005)\citenamefont{Ngai, Casalini,
  Capaccioli, Paluch, and Roland}}]{nga05}
\bibinfo{author}{\bibfnamefont{K.~L.} \bibnamefont{Ngai}},
  \bibinfo{author}{\bibfnamefont{R.}~\bibnamefont{Casalini}},
  \bibinfo{author}{\bibfnamefont{S.}~\bibnamefont{Capaccioli}},
  \bibinfo{author}{\bibfnamefont{M.}~\bibnamefont{Paluch}}, \bibnamefont{and}
  \bibinfo{author}{\bibfnamefont{C.~M.} \bibnamefont{Roland}},
  \bibinfo{journal}{J. Phys. Chem. B} \textbf{\bibinfo{volume}{109}},
  \bibinfo{pages}{17356} (\bibinfo{year}{2005}).

\bibitem[{\citenamefont{Coslovich and Roland}(2008)}]{cos08}
\bibinfo{author}{\bibfnamefont{D.}~\bibnamefont{Coslovich}} \bibnamefont{and}
  \bibinfo{author}{\bibfnamefont{C.~M.} \bibnamefont{Roland}},
  \bibinfo{journal}{J. Phys. Chem. B} \textbf{\bibinfo{volume}{112}},
  \bibinfo{pages}{1329} (\bibinfo{year}{2008}).

\bibitem[{\citenamefont{Coslovich and Roland}(2009)}]{cos09a}
\bibinfo{author}{\bibfnamefont{D.}~\bibnamefont{Coslovich}} \bibnamefont{and}
  \bibinfo{author}{\bibfnamefont{C.~M.} \bibnamefont{Roland}},
  \bibinfo{journal}{J. Chem. Phys.} \textbf{\bibinfo{volume}{130}},
  \bibinfo{pages}{014508} (\bibinfo{year}{2009}).

\bibitem[{\citenamefont{Roland}(2010)}]{rol10}
\bibinfo{author}{\bibfnamefont{C.~M.} \bibnamefont{Roland}},
  \bibinfo{journal}{Macromolecules} \textbf{\bibinfo{volume}{43}},
  \bibinfo{pages}{7875} (\bibinfo{year}{2010}).

\bibitem[{\citenamefont{Gundermann et~al.}(2011)\citenamefont{Gundermann,
  Pedersen, Hecksher, Bailey, Jakobsen, Christensen, Olsen, Schr{\o}der,
  Fragiadakis, Casalini et~al.}}]{gammagamma}
\bibinfo{author}{\bibfnamefont{D.}~\bibnamefont{Gundermann}},
  \bibinfo{author}{\bibfnamefont{U.~R.} \bibnamefont{Pedersen}},
  \bibinfo{author}{\bibfnamefont{T.}~\bibnamefont{Hecksher}},
  \bibinfo{author}{\bibfnamefont{N.~P.} \bibnamefont{Bailey}},
  \bibinfo{author}{\bibfnamefont{B.}~\bibnamefont{Jakobsen}},
  \bibinfo{author}{\bibfnamefont{T.}~\bibnamefont{Christensen}},
  \bibinfo{author}{\bibfnamefont{N.~B.} \bibnamefont{Olsen}},
  \bibinfo{author}{\bibfnamefont{T.~B.} \bibnamefont{Schr{\o}der}},
  \bibinfo{author}{\bibfnamefont{D.}~\bibnamefont{Fragiadakis}},
  \bibinfo{author}{\bibfnamefont{R.}~\bibnamefont{Casalini}},
  \bibnamefont{et~al.}, \bibinfo{journal}{Nature Phys.}
  \textbf{\bibinfo{volume}{7}}, \bibinfo{pages}{816} (\bibinfo{year}{2011}).

\bibitem[{\citenamefont{Ingebrigtsen
  et~al.}(2012{\natexlab{b}})\citenamefont{Ingebrigtsen, B{\o}hling,
  Schr{\o}der, and Dyre}}]{ing12a}
\bibinfo{author}{\bibfnamefont{T.~S.} \bibnamefont{Ingebrigtsen}},
  \bibinfo{author}{\bibfnamefont{L.}~\bibnamefont{B{\o}hling}},
  \bibinfo{author}{\bibfnamefont{T.~B.} \bibnamefont{Schr{\o}der}},
  \bibnamefont{and} \bibinfo{author}{\bibfnamefont{J.~C.} \bibnamefont{Dyre}},
  \bibinfo{journal}{J. Chem. Phys.} \textbf{\bibinfo{volume}{136}},
  \bibinfo{pages}{061102} (\bibinfo{year}{2012}{\natexlab{b}}).

\bibitem[{\citenamefont{Ashcroft and Lekner}(1966)}]{ash66}
\bibinfo{author}{\bibfnamefont{N.~W.} \bibnamefont{Ashcroft}} \bibnamefont{and}
  \bibinfo{author}{\bibfnamefont{J.}~\bibnamefont{Lekner}},
  \bibinfo{journal}{Phys. Rev.} \textbf{\bibinfo{volume}{145}},
  \bibinfo{pages}{83} (\bibinfo{year}{1966}).

\bibitem[{\citenamefont{Hansen and Verlet}(1969)}]{han69}
\bibinfo{author}{\bibfnamefont{J.-P.} \bibnamefont{Hansen}} \bibnamefont{and}
  \bibinfo{author}{\bibfnamefont{L.}~\bibnamefont{Verlet}},
  \bibinfo{journal}{Phys. Rev.} \textbf{\bibinfo{volume}{184}},
  \bibinfo{pages}{151.161} (\bibinfo{year}{1969}).

\bibitem[{\citenamefont{Grover}(1971)}]{gro71}
\bibinfo{author}{\bibfnamefont{R.}~\bibnamefont{Grover}}, \bibinfo{journal}{J.
  Chem. Phys.} \textbf{\bibinfo{volume}{55}}, \bibinfo{pages}{3435}
  (\bibinfo{year}{1971}).

\bibitem[{\citenamefont{Hoover et~al.}(1971)\citenamefont{Hoover, Gray, and
  Johnson}}]{hoo71}
\bibinfo{author}{\bibfnamefont{W.~G.} \bibnamefont{Hoover}},
  \bibinfo{author}{\bibfnamefont{S.~G.} \bibnamefont{Gray}}, \bibnamefont{and}
  \bibinfo{author}{\bibfnamefont{K.~W.} \bibnamefont{Johnson}},
  \bibinfo{journal}{J. Chem} \textbf{\bibinfo{volume}{55}},
  \bibinfo{pages}{1128} (\bibinfo{year}{1971}).

\bibitem[{\citenamefont{Hiwatari et~al.}(1974)\citenamefont{Hiwatari, Matsuda,
  Ogawa, Ogita, and Ueda}}]{hiw74}
\bibinfo{author}{\bibfnamefont{Y.}~\bibnamefont{Hiwatari}},
  \bibinfo{author}{\bibfnamefont{H.}~\bibnamefont{Matsuda}},
  \bibinfo{author}{\bibfnamefont{T.}~\bibnamefont{Ogawa}},
  \bibinfo{author}{\bibfnamefont{N.}~\bibnamefont{Ogita}}, \bibnamefont{and}
  \bibinfo{author}{\bibfnamefont{A.}~\bibnamefont{Ueda}},
  \bibinfo{journal}{Prog. Theor. Phys.} \textbf{\bibinfo{volume}{52}},
  \bibinfo{pages}{1105} (\bibinfo{year}{1974}).

\bibitem[{\citenamefont{Stishov}(1975)}]{sti75}
\bibinfo{author}{\bibfnamefont{S.~M.} \bibnamefont{Stishov}},
  \bibinfo{journal}{Sov. Phys. Usp.} \textbf{\bibinfo{volume}{17}},
  \bibinfo{pages}{625} (\bibinfo{year}{1975}).

\bibitem[{\citenamefont{Rosenfeld}(1976)}]{ros76}
\bibinfo{author}{\bibfnamefont{Y.}~\bibnamefont{Rosenfeld}},
  \bibinfo{journal}{Mol. Phys.} \textbf{\bibinfo{volume}{32}},
  \bibinfo{pages}{963} (\bibinfo{year}{1976}).

\bibitem[{\citenamefont{Kang et~al.}(1985)\citenamefont{Kang, Lee, Ree, and
  Ree}}]{kan85}
\bibinfo{author}{\bibfnamefont{H.~S.} \bibnamefont{Kang}},
  \bibinfo{author}{\bibfnamefont{S.~C.} \bibnamefont{Lee}},
  \bibinfo{author}{\bibfnamefont{T.}~\bibnamefont{Ree}}, \bibnamefont{and}
  \bibinfo{author}{\bibfnamefont{F.~H.} \bibnamefont{Ree}},
  \bibinfo{journal}{J. Chem. Phys.} \textbf{\bibinfo{volume}{82}},
  \bibinfo{pages}{414} (\bibinfo{year}{1985}).

\bibitem[{\citenamefont{Indrani and Ramaswamy}(1994)}]{ind94}
\bibinfo{author}{\bibfnamefont{A.~V.} \bibnamefont{Indrani}} \bibnamefont{and}
  \bibinfo{author}{\bibfnamefont{S.}~\bibnamefont{Ramaswamy}},
  \bibinfo{journal}{Phys. Rev. Lett} \textbf{\bibinfo{volume}{73}},
  \bibinfo{pages}{360} (\bibinfo{year}{1994}).

\bibitem[{\citenamefont{Branka and Heyes}(2006)}]{bra06}
\bibinfo{author}{\bibfnamefont{A.~C.} \bibnamefont{Branka}} \bibnamefont{and}
  \bibinfo{author}{\bibfnamefont{D.~M.} \bibnamefont{Heyes}},
  \bibinfo{journal}{Phys. Rev. E} \textbf{\bibinfo{volume}{74}},
  \bibinfo{pages}{031202} (\bibinfo{year}{2006}).

\bibitem[{\citenamefont{Heyes and Branka}(2007)}]{hey07}
\bibinfo{author}{\bibfnamefont{D.~M.} \bibnamefont{Heyes}} \bibnamefont{and}
  \bibinfo{author}{\bibfnamefont{A.~C.} \bibnamefont{Branka}},
  \bibinfo{journal}{Phys. Chem. Chem. Phys.} \textbf{\bibinfo{volume}{9}},
  \bibinfo{pages}{5570} (\bibinfo{year}{2007}).

\bibitem[{\citenamefont{Heyes and Branka}(2008)}]{hey08}
\bibinfo{author}{\bibfnamefont{D.~M.} \bibnamefont{Heyes}} \bibnamefont{and}
  \bibinfo{author}{\bibfnamefont{A.~C.} \bibnamefont{Branka}},
  \bibinfo{journal}{Phys. Chem. Chem. Phys.} \textbf{\bibinfo{volume}{10}},
  \bibinfo{pages}{4036} (\bibinfo{year}{2008}).

\bibitem[{\citenamefont{de~J.~Guevara-Rodriguez and
  Medina-Noyola}(2003)}]{gue03}
\bibinfo{author}{\bibfnamefont{F.}~\bibnamefont{de~J.~Guevara-Rodriguez}}
  \bibnamefont{and}
  \bibinfo{author}{\bibfnamefont{M.}~\bibnamefont{Medina-Noyola}},
  \bibinfo{journal}{Phys. Rev. E} \textbf{\bibinfo{volume}{68}},
  \bibinfo{pages}{011405} (\bibinfo{year}{2003}).

\bibitem[{\citenamefont{Lopez-Flores et~al.}(2011)\citenamefont{Lopez-Flores,
  Mendoza-Mendez, Sanchez-Diaz, Perez-Angel, Chavez-Paez, Viscarra-Rendion, and
  Medina-Noyola}}]{lop11a}
\bibinfo{author}{\bibfnamefont{L.}~\bibnamefont{Lopez-Flores}},
  \bibinfo{author}{\bibfnamefont{P.}~\bibnamefont{Mendoza-Mendez}},
  \bibinfo{author}{\bibfnamefont{L.~E.} \bibnamefont{Sanchez-Diaz}},
  \bibinfo{author}{\bibfnamefont{G.}~\bibnamefont{Perez-Angel}},
  \bibinfo{author}{\bibfnamefont{M.}~\bibnamefont{Chavez-Paez}},
  \bibinfo{author}{\bibfnamefont{A.}~\bibnamefont{Viscarra-Rendion}},
  \bibnamefont{and}
  \bibinfo{author}{\bibfnamefont{M.}~\bibnamefont{Medina-Noyola}},
  \bibinfo{journal}{arXiv:1106.2475}  (\bibinfo{year}{2011}).

\bibitem[{\citenamefont{Young and Andersen}(2003)}]{you03}
\bibinfo{author}{\bibfnamefont{T.}~\bibnamefont{Young}} \bibnamefont{and}
  \bibinfo{author}{\bibfnamefont{H.~C.} \bibnamefont{Andersen}},
  \bibinfo{journal}{J. Chem. Phys.} \textbf{\bibinfo{volume}{118}},
  \bibinfo{pages}{3447} (\bibinfo{year}{2003}).

\bibitem[{\citenamefont{Young and Andersen}(2005)}]{you05}
\bibinfo{author}{\bibfnamefont{T.}~\bibnamefont{Young}} \bibnamefont{and}
  \bibinfo{author}{\bibfnamefont{H.~C.} \bibnamefont{Andersen}},
  \bibinfo{journal}{J. Phys. Chem. B} \textbf{\bibinfo{volume}{109}},
  \bibinfo{pages}{2985} (\bibinfo{year}{2005}).

\bibitem[{\citenamefont{Pond et~al.}(2011{\natexlab{a}})\citenamefont{Pond,
  Errington, and Truskett}}]{pon11a}
\bibinfo{author}{\bibfnamefont{M.~J.} \bibnamefont{Pond}},
  \bibinfo{author}{\bibfnamefont{J.~R.} \bibnamefont{Errington}},
  \bibnamefont{and} \bibinfo{author}{\bibfnamefont{T.~M.}
  \bibnamefont{Truskett}}, \bibinfo{journal}{Soft Matter}
  \textbf{\bibinfo{volume}{7}}, \bibinfo{pages}{9859}
  (\bibinfo{year}{2011}{\natexlab{a}}).

\bibitem[{\citenamefont{Schmiedeberg et~al.}(2011)\citenamefont{Schmiedeberg,
  Haxton, Nagel, and Liu}}]{sch11}
\bibinfo{author}{\bibfnamefont{M.}~\bibnamefont{Schmiedeberg}},
  \bibinfo{author}{\bibfnamefont{T.~K.} \bibnamefont{Haxton}},
  \bibinfo{author}{\bibfnamefont{S.~R.} \bibnamefont{Nagel}}, \bibnamefont{and}
  \bibinfo{author}{\bibfnamefont{A.~J.} \bibnamefont{Liu}},
  \bibinfo{journal}{EPL} \textbf{\bibinfo{volume}{96}}, \bibinfo{pages}{36010}
  (\bibinfo{year}{2011}).

\bibitem[{\citenamefont{Noro and Frenkel}(2000)}]{nor00}
\bibinfo{author}{\bibfnamefont{M.~G.} \bibnamefont{Noro}} \bibnamefont{and}
  \bibinfo{author}{\bibfnamefont{D.}~\bibnamefont{Frenkel}},
  \bibinfo{journal}{J. Chem. Phys.} \textbf{\bibinfo{volume}{113}},
  \bibinfo{pages}{2941} (\bibinfo{year}{2000}).

\bibitem[{\citenamefont{Speedy}(2003)}]{spe03}
\bibinfo{author}{\bibfnamefont{R.~J.} \bibnamefont{Speedy}},
  \bibinfo{journal}{J. Phys.: Condens. Matter} \textbf{\bibinfo{volume}{15}},
  \bibinfo{pages}{S1243} (\bibinfo{year}{2003}).

\bibitem[{\citenamefont{Sheng and Ma}(2004)}]{she04}
\bibinfo{author}{\bibfnamefont{H.~W.} \bibnamefont{Sheng}} \bibnamefont{and}
  \bibinfo{author}{\bibfnamefont{E.}~\bibnamefont{Ma}}, \bibinfo{journal}{Phys.
  Rev. E} \textbf{\bibinfo{volume}{69}}, \bibinfo{pages}{062202}
  (\bibinfo{year}{2004}).

\bibitem[{\citenamefont{Heyes and Branka}(2005)}]{hey05}
\bibinfo{author}{\bibfnamefont{D.~M.} \bibnamefont{Heyes}} \bibnamefont{and}
  \bibinfo{author}{\bibfnamefont{A.~C.} \bibnamefont{Branka}},
  \bibinfo{journal}{J. Chem. Phys.} \textbf{\bibinfo{volume}{122}},
  \bibinfo{pages}{234504} (\bibinfo{year}{2005}).

\bibitem[{\citenamefont{Scopigno et~al.}(2005)\citenamefont{Scopigno, {Di
  Leonardo}, Comez, Baron, Fioretto, and Ruocco}}]{sco05}
\bibinfo{author}{\bibfnamefont{T.}~\bibnamefont{Scopigno}},
  \bibinfo{author}{\bibfnamefont{R.}~\bibnamefont{{Di Leonardo}}},
  \bibinfo{author}{\bibfnamefont{L.}~\bibnamefont{Comez}},
  \bibinfo{author}{\bibfnamefont{A.~Q.~R.} \bibnamefont{Baron}},
  \bibinfo{author}{\bibfnamefont{D.}~\bibnamefont{Fioretto}}, \bibnamefont{and}
  \bibinfo{author}{\bibfnamefont{G.}~\bibnamefont{Ruocco}},
  \bibinfo{journal}{Phys. Rev. Lett.} \textbf{\bibinfo{volume}{94}},
  \bibinfo{pages}{155301} (\bibinfo{year}{2005}).

\bibitem[{\citenamefont{Orea et~al.}(2008)\citenamefont{Orea, Reyes-Mercado,
  and Duda}}]{ore08}
\bibinfo{author}{\bibfnamefont{P.}~\bibnamefont{Orea}},
  \bibinfo{author}{\bibfnamefont{Y.}~\bibnamefont{Reyes-Mercado}},
  \bibnamefont{and} \bibinfo{author}{\bibfnamefont{Y.}~\bibnamefont{Duda}},
  \bibinfo{journal}{Phys. Lett. A} \textbf{\bibinfo{volume}{372}},
  \bibinfo{pages}{7024} (\bibinfo{year}{2008}).

\bibitem[{\citenamefont{Heyes and Branka}(2009)}]{hey09}
\bibinfo{author}{\bibfnamefont{D.~M.} \bibnamefont{Heyes}} \bibnamefont{and}
  \bibinfo{author}{\bibfnamefont{A.~C.} \bibnamefont{Branka}},
  \bibinfo{journal}{Mol. Phys.} \textbf{\bibinfo{volume}{107}},
  \bibinfo{pages}{309} (\bibinfo{year}{2009}).

\bibitem[{\citenamefont{Lange et~al.}(2009)\citenamefont{Lange, Caballero,
  Puertas, and Fuchs}}]{lan09}
\bibinfo{author}{\bibfnamefont{E.}~\bibnamefont{Lange}},
  \bibinfo{author}{\bibfnamefont{J.~B.} \bibnamefont{Caballero}},
  \bibinfo{author}{\bibfnamefont{A.~M.} \bibnamefont{Puertas}},
  \bibnamefont{and} \bibinfo{author}{\bibfnamefont{M.}~\bibnamefont{Fuchs}},
  \bibinfo{journal}{J. Chem. Phys.} \textbf{\bibinfo{volume}{130}},
  \bibinfo{pages}{174903} (\bibinfo{year}{2009}).

\bibitem[{\citenamefont{Ramirez-Gonzalez and Medina-Noyola}(2009)}]{ram09}
\bibinfo{author}{\bibfnamefont{P.~E.} \bibnamefont{Ramirez-Gonzalez}}
  \bibnamefont{and}
  \bibinfo{author}{\bibfnamefont{M.}~\bibnamefont{Medina-Noyola}},
  \bibinfo{journal}{J. Phys.: Condens. Matter} \textbf{\bibinfo{volume}{21}},
  \bibinfo{pages}{075101} (\bibinfo{year}{2009}).

\bibitem[{\citenamefont{Ramirez-Gonzalez
  et~al.}(2011)\citenamefont{Ramirez-Gonzalez, Lopez-Flores, Acuna-Campa, and
  Medina-Noyola}}]{ram11}
\bibinfo{author}{\bibfnamefont{P.~E.} \bibnamefont{Ramirez-Gonzalez}},
  \bibinfo{author}{\bibfnamefont{L.}~\bibnamefont{Lopez-Flores}},
  \bibinfo{author}{\bibfnamefont{H.}~\bibnamefont{Acuna-Campa}},
  \bibnamefont{and}
  \bibinfo{author}{\bibfnamefont{M.}~\bibnamefont{Medina-Noyola}},
  \bibinfo{journal}{Phys. Rev. Lett.} \textbf{\bibinfo{volume}{107}},
  \bibinfo{pages}{155701} (\bibinfo{year}{2011}).

\bibitem[{\citenamefont{Truskett et~al.}(2000)\citenamefont{Truskett, Torquato,
  and Debenedetti}}]{tru00}
\bibinfo{author}{\bibfnamefont{T.~M.} \bibnamefont{Truskett}},
  \bibinfo{author}{\bibfnamefont{S.}~\bibnamefont{Torquato}}, \bibnamefont{and}
  \bibinfo{author}{\bibfnamefont{P.~G.} \bibnamefont{Debenedetti}},
  \bibinfo{journal}{Phys. Rev. E} \textbf{\bibinfo{volume}{62}},
  \bibinfo{pages}{993} (\bibinfo{year}{2000}).

\bibitem[{\citenamefont{Errington et~al.}(2003)\citenamefont{Errington,
  Debenedetti, and Torquato}}]{err03}
\bibinfo{author}{\bibfnamefont{J.~R.} \bibnamefont{Errington}},
  \bibinfo{author}{\bibfnamefont{P.~G.} \bibnamefont{Debenedetti}},
  \bibnamefont{and} \bibinfo{author}{\bibfnamefont{S.}~\bibnamefont{Torquato}},
  \bibinfo{journal}{J. Chem. Phys.} \textbf{\bibinfo{volume}{118}},
  \bibinfo{pages}{2256} (\bibinfo{year}{2003}).

\bibitem[{\citenamefont{Chakraborty and Chakravarty}(2007)}]{cha07}
\bibinfo{author}{\bibfnamefont{S.~N.} \bibnamefont{Chakraborty}}
  \bibnamefont{and}
  \bibinfo{author}{\bibfnamefont{C.}~\bibnamefont{Chakravarty}},
  \bibinfo{journal}{Phys. Rev. E} \textbf{\bibinfo{volume}{76}},
  \bibinfo{pages}{011201} (\bibinfo{year}{2007}).

\bibitem[{\citenamefont{Rosenfeld}(1977)}]{ros77}
\bibinfo{author}{\bibfnamefont{Y.}~\bibnamefont{Rosenfeld}},
  \bibinfo{journal}{Phys. Rev. A} \textbf{\bibinfo{volume}{15}},
  \bibinfo{pages}{2545} (\bibinfo{year}{1977}).

\bibitem[{\citenamefont{Rosenfeld}(1999)}]{ros99}
\bibinfo{author}{\bibfnamefont{Y.}~\bibnamefont{Rosenfeld}},
  \bibinfo{journal}{J. Phys.: Condens. Matter} \textbf{\bibinfo{volume}{11}},
  \bibinfo{pages}{5415} (\bibinfo{year}{1999}).

\bibitem[{\citenamefont{Pond et~al.}(2011{\natexlab{b}})\citenamefont{Pond,
  Errington, and Truskett}}]{pon11}
\bibinfo{author}{\bibfnamefont{M.~J.} \bibnamefont{Pond}},
  \bibinfo{author}{\bibfnamefont{J.~R.} \bibnamefont{Errington}},
  \bibnamefont{and} \bibinfo{author}{\bibfnamefont{T.~M.}
  \bibnamefont{Truskett}}, \bibinfo{journal}{J. Chem. Phys.}
  \textbf{\bibinfo{volume}{134}}, \bibinfo{pages}{081101}
  (\bibinfo{year}{2011}{\natexlab{b}}).

\bibitem[{\citenamefont{Singh et~al.}(2012)\citenamefont{Singh, Agarwal,
  Dhabal, and Chakravarty}}]{sin12}
\bibinfo{author}{\bibfnamefont{M.}~\bibnamefont{Singh}},
  \bibinfo{author}{\bibfnamefont{M.}~\bibnamefont{Agarwal}},
  \bibinfo{author}{\bibfnamefont{D.}~\bibnamefont{Dhabal}}, \bibnamefont{and}
  \bibinfo{author}{\bibfnamefont{C.}~\bibnamefont{Chakravarty}},
  \bibinfo{journal}{J. Chem. Phys.} \textbf{\bibinfo{volume}{137}},
  \bibinfo{pages}{024508} (\bibinfo{year}{2012}).

\bibitem[{\citenamefont{Grover et~al.}(1985)\citenamefont{Grover, Hoover, and
  Moran}}]{gro85}
\bibinfo{author}{\bibfnamefont{R.}~\bibnamefont{Grover}},
  \bibinfo{author}{\bibfnamefont{W.~G.} \bibnamefont{Hoover}},
  \bibnamefont{and} \bibinfo{author}{\bibfnamefont{B.}~\bibnamefont{Moran}},
  \bibinfo{journal}{J. Chem. Phys.} \textbf{\bibinfo{volume}{83}},
  \bibinfo{pages}{1255} (\bibinfo{year}{1985}).

\bibitem[{\citenamefont{Gilvarry}(1956)}]{gil56}
\bibinfo{author}{\bibfnamefont{J.~J.} \bibnamefont{Gilvarry}},
  \bibinfo{journal}{Phys. Rev.} \textbf{\bibinfo{volume}{102}},
  \bibinfo{pages}{308} (\bibinfo{year}{1956}).

\bibitem[{\citenamefont{Ubbelohde}(1965)}]{ubb65}
\bibinfo{author}{\bibfnamefont{A.~R.} \bibnamefont{Ubbelohde}},
  \emph{\bibinfo{title}{Melting and Crystal Structure}}
  (\bibinfo{publisher}{Clarendon (London)}, \bibinfo{year}{1965}).

\bibitem[{\citenamefont{Ross}(1969)}]{ros69}
\bibinfo{author}{\bibfnamefont{M.}~\bibnamefont{Ross}}, \bibinfo{journal}{Phys.
  Rev.} \textbf{\bibinfo{volume}{184}}, \bibinfo{pages}{233}
  (\bibinfo{year}{1969}).

\bibitem[{\citenamefont{Saija et~al.}(2006)\citenamefont{Saija, Prestipino, and
  Giaquinta}}]{sai06}
\bibinfo{author}{\bibfnamefont{F.}~\bibnamefont{Saija}},
  \bibinfo{author}{\bibfnamefont{S.}~\bibnamefont{Prestipino}},
  \bibnamefont{and} \bibinfo{author}{\bibfnamefont{P.~V.}
  \bibnamefont{Giaquinta}}, \bibinfo{journal}{J. Chem. Phys.}
  \textbf{\bibinfo{volume}{124}}, \bibinfo{pages}{244504}
  (\bibinfo{year}{2006}).

\bibitem[{\citenamefont{Malescio et~al.}(2000)\citenamefont{Malescio,
  Giaquinta, and Rosenfeld}}]{mal00}
\bibinfo{author}{\bibfnamefont{G.}~\bibnamefont{Malescio}},
  \bibinfo{author}{\bibfnamefont{P.~V.} \bibnamefont{Giaquinta}},
  \bibnamefont{and}
  \bibinfo{author}{\bibfnamefont{Y.}~\bibnamefont{Rosenfeld}},
  \bibinfo{journal}{Phys. Rev. E} \textbf{\bibinfo{volume}{61}},
  \bibinfo{pages}{4090} (\bibinfo{year}{2000}).

\bibitem[{\citenamefont{Toxvaerd}()}]{st}
\bibinfo{author}{\bibfnamefont{S.}~\bibnamefont{Toxvaerd}},
  \bibinfo{note}{{unpublished}}.

\bibitem[{\citenamefont{Wallace}(2002)}]{wallace}
\bibinfo{author}{\bibfnamefont{D.~C.} \bibnamefont{Wallace}},
  \emph{\bibinfo{title}{Statistical Physics of Crystals and Liquids}}
  (\bibinfo{publisher}{World Scientific, Singapore}, \bibinfo{year}{2002}).

\bibitem[{\citenamefont{Bolmatov et~al.}(2012)\citenamefont{Bolmatov, Brazhkin,
  and Tracenko}}]{bol12}
\bibinfo{author}{\bibfnamefont{D.}~\bibnamefont{Bolmatov}},
  \bibinfo{author}{\bibfnamefont{V.~V.} \bibnamefont{Brazhkin}},
  \bibnamefont{and} \bibinfo{author}{\bibfnamefont{K.}~\bibnamefont{Tracenko}},
  \bibinfo{journal}{Sci. Rep.} \textbf{\bibinfo{volume}{2}},
  \bibinfo{pages}{421} (\bibinfo{year}{2012}).

\bibitem[{\citenamefont{Andrade}(1934)}]{and34}
\bibinfo{author}{\bibfnamefont{E.~N.~D.} \bibnamefont{Andrade}},
  \bibinfo{journal}{Phil. Mag.} \textbf{\bibinfo{volume}{17}},
  \bibinfo{pages}{497} (\bibinfo{year}{1934}).

\bibitem[{\citenamefont{Kaptay}(2005)}]{kap05}
\bibinfo{author}{\bibfnamefont{G.}~\bibnamefont{Kaptay}}, \bibinfo{journal}{Z.
  Metallkd.} \textbf{\bibinfo{volume}{96}}, \bibinfo{pages}{1}
  (\bibinfo{year}{2005}).

\bibitem[{\citenamefont{Raveche et~al.}(1974)\citenamefont{Raveche, Mountain,
  and Streett}}]{rav74}
\bibinfo{author}{\bibfnamefont{H.~J.} \bibnamefont{Raveche}},
  \bibinfo{author}{\bibfnamefont{R.~D.} \bibnamefont{Mountain}},
  \bibnamefont{and} \bibinfo{author}{\bibfnamefont{W.~B.}
  \bibnamefont{Streett}}, \bibinfo{journal}{J. Chem. Phys.}
  \textbf{\bibinfo{volume}{61}}, \bibinfo{pages}{1970} (\bibinfo{year}{1974}).

\bibitem[{\citenamefont{Saija et~al.}({2001})\citenamefont{Saija, Prestipino,
  and Giaquinta}}]{sai01}
\bibinfo{author}{\bibfnamefont{F.}~\bibnamefont{Saija}},
  \bibinfo{author}{\bibfnamefont{S.}~\bibnamefont{Prestipino}},
  \bibnamefont{and} \bibinfo{author}{\bibfnamefont{P.~V.}
  \bibnamefont{Giaquinta}}, \bibinfo{journal}{J. Chem. Phys.}
  \textbf{\bibinfo{volume}{{115}}}, \bibinfo{pages}{7586}
  (\bibinfo{year}{{2001}}).

\bibitem[{\citenamefont{Tallon}(1980)}]{tal80}
\bibinfo{author}{\bibfnamefont{J.~L.} \bibnamefont{Tallon}},
  \bibinfo{journal}{Phys. Lett. A} \textbf{\bibinfo{volume}{76}},
  \bibinfo{pages}{139} (\bibinfo{year}{1980}).

\bibitem[{\citenamefont{Rosenfeld}(1982)}]{ros82}
\bibinfo{author}{\bibfnamefont{Y.}~\bibnamefont{Rosenfeld}},
  \bibinfo{journal}{Phys. Rev. A} \textbf{\bibinfo{volume}{26}},
  \bibinfo{pages}{3633} (\bibinfo{year}{1982}).

\bibitem[{\citenamefont{Khrapak and Morfill}(2011)}]{khr11}
\bibinfo{author}{\bibfnamefont{S.~A.} \bibnamefont{Khrapak}} \bibnamefont{and}
  \bibinfo{author}{\bibfnamefont{G.~E.} \bibnamefont{Morfill}},
  \bibinfo{journal}{J. Chem. Phys.} \textbf{\bibinfo{volume}{134}},
  \bibinfo{pages}{094108} (\bibinfo{year}{2011}).

\bibitem[{\citenamefont{Alba-Simionesco
  et~al.}(2002)\citenamefont{Alba-Simionesco, Kivelson, and Tarjus}}]{alb02}
\bibinfo{author}{\bibfnamefont{C.}~\bibnamefont{Alba-Simionesco}},
  \bibinfo{author}{\bibfnamefont{D.}~\bibnamefont{Kivelson}}, \bibnamefont{and}
  \bibinfo{author}{\bibfnamefont{G.}~\bibnamefont{Tarjus}},
  \bibinfo{journal}{J. Chem. Phys.} \textbf{\bibinfo{volume}{116}},
  \bibinfo{pages}{5033} (\bibinfo{year}{2002}).

\bibitem[{\citenamefont{Sciortino et~al.}(1999)\citenamefont{Sciortino, Kob,
  and Tartaglia}}]{sci99}
\bibinfo{author}{\bibfnamefont{F.}~\bibnamefont{Sciortino}},
  \bibinfo{author}{\bibfnamefont{W.}~\bibnamefont{Kob}}, \bibnamefont{and}
  \bibinfo{author}{\bibfnamefont{P.}~\bibnamefont{Tartaglia}},
  \bibinfo{journal}{Phys. Rev. Lett.} \textbf{\bibinfo{volume}{83}},
  \bibinfo{pages}{3214} (\bibinfo{year}{1999}).

\bibitem[{\citenamefont{Doliwa and Heuer}(2003)}]{dol03}
\bibinfo{author}{\bibfnamefont{B.}~\bibnamefont{Doliwa}} \bibnamefont{and}
  \bibinfo{author}{\bibfnamefont{A.}~\bibnamefont{Heuer}}, \bibinfo{journal}{J.
  Phys.: Condens. Matter} \textbf{\bibinfo{volume}{15}}, \bibinfo{pages}{S849}
  (\bibinfo{year}{2003}).

\bibitem[{\citenamefont{Yan et~al.}(2004)\citenamefont{Yan, Jain, and {de
  Pablo}}}]{yan04}
\bibinfo{author}{\bibfnamefont{Q.}~\bibnamefont{Yan}},
  \bibinfo{author}{\bibfnamefont{T.~S.} \bibnamefont{Jain}}, \bibnamefont{and}
  \bibinfo{author}{\bibfnamefont{J.~J.} \bibnamefont{{de Pablo}}},
  \bibinfo{journal}{Phys. Rev. Lett.} \textbf{\bibinfo{volume}{92}},
  \bibinfo{pages}{235701} (\bibinfo{year}{2004}).

\bibitem[{\citenamefont{Gebremichael et~al.}({2005})\citenamefont{Gebremichael,
  Vogel, Bergroth, Starr, and Glotzer}}]{geb05}
\bibinfo{author}{\bibfnamefont{Y.}~\bibnamefont{Gebremichael}},
  \bibinfo{author}{\bibfnamefont{M.}~\bibnamefont{Vogel}},
  \bibinfo{author}{\bibfnamefont{M.~N.~J.} \bibnamefont{Bergroth}},
  \bibinfo{author}{\bibfnamefont{F.~W.} \bibnamefont{Starr}}, \bibnamefont{and}
  \bibinfo{author}{\bibfnamefont{S.~C.} \bibnamefont{Glotzer}},
  \bibinfo{journal}{J. Phys. Chem. B} \textbf{\bibinfo{volume}{{109}}},
  \bibinfo{pages}{15068} (\bibinfo{year}{{2005}}).

\bibitem[{\citenamefont{Pedersen et~al.}(2010)\citenamefont{Pedersen,
  Schr{\o}der, and Dyre}}]{ped10}
\bibinfo{author}{\bibfnamefont{U.~R.} \bibnamefont{Pedersen}},
  \bibinfo{author}{\bibfnamefont{T.~B.} \bibnamefont{Schr{\o}der}},
  \bibnamefont{and} \bibinfo{author}{\bibfnamefont{J.~C.} \bibnamefont{Dyre}},
  \bibinfo{journal}{Phys. Rev. Lett.} \textbf{\bibinfo{volume}{105}},
  \bibinfo{pages}{157801} (\bibinfo{year}{2010}).

\bibitem[{\citenamefont{Ingebrigtsen and Dyre}()}]{trond}
\bibinfo{author}{\bibfnamefont{T.~S.} \bibnamefont{Ingebrigtsen}}
  \bibnamefont{and} \bibinfo{author}{\bibfnamefont{J.~C.} \bibnamefont{Dyre}},
  \bibinfo{note}{{unpublished}}.

\bibitem[{\citenamefont{Angell}(1995)}]{ang95}
\bibinfo{author}{\bibfnamefont{C.~A.} \bibnamefont{Angell}},
  \bibinfo{journal}{Science} \textbf{\bibinfo{volume}{267}},
  \bibinfo{pages}{1924} (\bibinfo{year}{1995}).

\bibitem[{\citenamefont{Johari}(1974)}]{joh74}
\bibinfo{author}{\bibfnamefont{G.~P.} \bibnamefont{Johari}},
  \bibinfo{journal}{J. Chem. Educ.} \textbf{\bibinfo{volume}{51}},
  \bibinfo{pages}{23} (\bibinfo{year}{1974}).

\bibitem[{\citenamefont{Dyre}(2006)}]{dyr06}
\bibinfo{author}{\bibfnamefont{J.~C.} \bibnamefont{Dyre}},
  \bibinfo{journal}{Rev. Mod. Phys.} \textbf{\bibinfo{volume}{78}},
  \bibinfo{pages}{953} (\bibinfo{year}{2006}).

\bibitem[{\citenamefont{Kivelson et~al.}(1996)\citenamefont{Kivelson, Tarjus,
  Zhao, and Kivelson}}]{kiv96}
\bibinfo{author}{\bibfnamefont{D.}~\bibnamefont{Kivelson}},
  \bibinfo{author}{\bibfnamefont{G.}~\bibnamefont{Tarjus}},
  \bibinfo{author}{\bibfnamefont{X.}~\bibnamefont{Zhao}}, \bibnamefont{and}
  \bibinfo{author}{\bibfnamefont{S.~A.} \bibnamefont{Kivelson}},
  \bibinfo{journal}{Phys. Rev. E} \textbf{\bibinfo{volume}{53}},
  \bibinfo{pages}{751} (\bibinfo{year}{1996}).

\bibitem[{\citenamefont{{De Michele} et~al.}(2004)\citenamefont{{De Michele},
  Sciortino, and Coniglio}}]{dem04}
\bibinfo{author}{\bibfnamefont{C.}~\bibnamefont{{De Michele}}},
  \bibinfo{author}{\bibfnamefont{F.}~\bibnamefont{Sciortino}},
  \bibnamefont{and} \bibinfo{author}{\bibfnamefont{A.}~\bibnamefont{Coniglio}},
  \bibinfo{journal}{J. Phys.: Condens. Matter} \textbf{\bibinfo{volume}{16}},
  \bibinfo{pages}{L489} (\bibinfo{year}{2004}).

\bibitem[{\citenamefont{Sengupta et~al.}(2011)\citenamefont{Sengupta,
  Vasconcelos, Affouard, and Sastry}}]{sen11}
\bibinfo{author}{\bibfnamefont{S.}~\bibnamefont{Sengupta}},
  \bibinfo{author}{\bibfnamefont{F.}~\bibnamefont{Vasconcelos}},
  \bibinfo{author}{\bibfnamefont{F.}~\bibnamefont{Affouard}}, \bibnamefont{and}
  \bibinfo{author}{\bibfnamefont{S.}~\bibnamefont{Sastry}},
  \bibinfo{journal}{J. Chem. Phys.} \textbf{\bibinfo{volume}{135}},
  \bibinfo{pages}{194503} (\bibinfo{year}{2011}).

\bibitem[{\citenamefont{Niss et~al.}(2007)\citenamefont{Niss, Dalle-Ferrier,
  Tarjus, and Alba-Simionesco}}]{nis07}
\bibinfo{author}{\bibfnamefont{K.}~\bibnamefont{Niss}},
  \bibinfo{author}{\bibfnamefont{C.}~\bibnamefont{Dalle-Ferrier}},
  \bibinfo{author}{\bibfnamefont{G.}~\bibnamefont{Tarjus}}, \bibnamefont{and}
  \bibinfo{author}{\bibfnamefont{C.}~\bibnamefont{Alba-Simionesco}},
  \bibinfo{journal}{J. Phys.: Condens. Matter} \textbf{\bibinfo{volume}{19}},
  \bibinfo{pages}{076102} (\bibinfo{year}{2007}).

\bibitem[{\citenamefont{Zwanzig}(1954)}]{zwa54}
\bibinfo{author}{\bibfnamefont{R.~W.} \bibnamefont{Zwanzig}},
  \bibinfo{journal}{J. Chem. Phys.} \textbf{\bibinfo{volume}{22}},
  \bibinfo{pages}{1420} (\bibinfo{year}{1954}).

\bibitem[{\citenamefont{Longuet-Higgins and Widom}(1964)}]{lon64}
\bibinfo{author}{\bibfnamefont{H.~C.} \bibnamefont{Longuet-Higgins}}
  \bibnamefont{and} \bibinfo{author}{\bibfnamefont{B.}~\bibnamefont{Widom}},
  \bibinfo{journal}{Mol. Phys.} \textbf{\bibinfo{volume}{8}},
  \bibinfo{pages}{549} (\bibinfo{year}{1964}).

\bibitem[{\citenamefont{Widom}(1967)}]{wid67}
\bibinfo{author}{\bibfnamefont{B.}~\bibnamefont{Widom}},
  \bibinfo{journal}{Science} \textbf{\bibinfo{volume}{157}},
  \bibinfo{pages}{375} (\bibinfo{year}{1967}).

\bibitem[{\citenamefont{Weeks et~al.}(1971)\citenamefont{Weeks, Chandler, and
  Andersen}}]{wca}
\bibinfo{author}{\bibfnamefont{J.~D.} \bibnamefont{Weeks}},
  \bibinfo{author}{\bibfnamefont{D.}~\bibnamefont{Chandler}}, \bibnamefont{and}
  \bibinfo{author}{\bibfnamefont{H.~C.} \bibnamefont{Andersen}},
  \bibinfo{journal}{J. Chem. Phys.} \textbf{\bibinfo{volume}{54}},
  \bibinfo{pages}{5237} (\bibinfo{year}{1971}).

\bibitem[{\citenamefont{Torquato and Stillinger}(2010)}]{tor10}
\bibinfo{author}{\bibfnamefont{S.}~\bibnamefont{Torquato}} \bibnamefont{and}
  \bibinfo{author}{\bibfnamefont{F.~H.} \bibnamefont{Stillinger}},
  \bibinfo{journal}{Rev. Mod. Phys.} \textbf{\bibinfo{volume}{82}},
  \bibinfo{pages}{2633} (\bibinfo{year}{2010}).

\bibitem[{\citenamefont{Brito and Wyart}(2009)}]{bri09}
\bibinfo{author}{\bibfnamefont{C.}~\bibnamefont{Brito}} \bibnamefont{and}
  \bibinfo{author}{\bibfnamefont{M.}~\bibnamefont{Wyart}}, \bibinfo{journal}{J.
  Chem. Phys.} \textbf{\bibinfo{volume}{131}}, \bibinfo{pages}{024504}
  (\bibinfo{year}{2009}).

\bibitem[{\citenamefont{Rosenfeld}(1998)}]{ros98a}
\bibinfo{author}{\bibfnamefont{Y.}~\bibnamefont{Rosenfeld}},
  \bibinfo{journal}{Mol. Phys.} \textbf{\bibinfo{volume}{94}},
  \bibinfo{pages}{929} (\bibinfo{year}{1998}).

\bibitem[{\citenamefont{Bestul and Chang}(1964)}]{bes64}
\bibinfo{author}{\bibfnamefont{A.~B.} \bibnamefont{Bestul}} \bibnamefont{and}
  \bibinfo{author}{\bibfnamefont{S.~S.} \bibnamefont{Chang}},
  \bibinfo{journal}{J. Chem. Phys.} \textbf{\bibinfo{volume}{40}},
  \bibinfo{pages}{3731} (\bibinfo{year}{1964}).

\bibitem[{\citenamefont{Adam and Gibbs}(1965)}]{ada65}
\bibinfo{author}{\bibfnamefont{G.}~\bibnamefont{Adam}} \bibnamefont{and}
  \bibinfo{author}{\bibfnamefont{J.~H.} \bibnamefont{Gibbs}},
  \bibinfo{journal}{J. Chem. Phys.} \textbf{\bibinfo{volume}{43}},
  \bibinfo{pages}{139} (\bibinfo{year}{1965}).

\bibitem[{\citenamefont{Dudowicz et~al.}(2008)\citenamefont{Dudowicz, Freed,
  and Douglas}}]{dud08}
\bibinfo{author}{\bibfnamefont{J.}~\bibnamefont{Dudowicz}},
  \bibinfo{author}{\bibfnamefont{K.~F.} \bibnamefont{Freed}}, \bibnamefont{and}
  \bibinfo{author}{\bibfnamefont{J.~F.} \bibnamefont{Douglas}},
  \bibinfo{journal}{Adv. Chem. Phys.} \textbf{\bibinfo{volume}{138}},
  \bibinfo{pages}{125} (\bibinfo{year}{2008}).

\bibitem[{\citenamefont{Dyre et~al.}(2009)\citenamefont{Dyre, Heckhsher, and
  Niss}}]{dyr09}
\bibinfo{author}{\bibfnamefont{J.~C.} \bibnamefont{Dyre}},
  \bibinfo{author}{\bibfnamefont{T.}~\bibnamefont{Heckhsher}},
  \bibnamefont{and} \bibinfo{author}{\bibfnamefont{K.}~\bibnamefont{Niss}},
  \bibinfo{journal}{J. Non-Cryst. Solids} \textbf{\bibinfo{volume}{355}},
  \bibinfo{pages}{624} (\bibinfo{year}{2009}).

\bibitem[{\citenamefont{Dzugutov}(1996)}]{dzu96}
\bibinfo{author}{\bibfnamefont{M.}~\bibnamefont{Dzugutov}},
  \bibinfo{journal}{Nature} \textbf{\bibinfo{volume}{381}},
  \bibinfo{pages}{139} (\bibinfo{year}{1996}).

\bibitem[{\citenamefont{Hicks}(1963)}]{hicks}
\bibinfo{author}{\bibfnamefont{N.~J.} \bibnamefont{Hicks}},
  \emph{\bibinfo{title}{Notes on Differential Geometry}}
  (\bibinfo{publisher}{van Nostrand Reinhold, New York}, \bibinfo{year}{1963}).

\bibitem[{\citenamefont{Dombrowski}(1968)}]{dom68}
\bibinfo{author}{\bibfnamefont{P.}~\bibnamefont{Dombrowski}},
  \bibinfo{journal}{Math. Nachr.} \textbf{\bibinfo{volume}{38}},
  \bibinfo{pages}{133} (\bibinfo{year}{1968}).

\bibitem[{\citenamefont{Gallot et~al.}(2004)\citenamefont{Gallot, Hulin, and
  Lafontaine}}]{riemannian}
\bibinfo{author}{\bibfnamefont{S.}~\bibnamefont{Gallot}},
  \bibinfo{author}{\bibfnamefont{D.}~\bibnamefont{Hulin}}, \bibnamefont{and}
  \bibinfo{author}{\bibfnamefont{J.}~\bibnamefont{Lafontaine}},
  \emph{\bibinfo{title}{Riemannian Geometry}} (\bibinfo{publisher}{Springer,
  Berlin}, \bibinfo{year}{2004}).

\bibitem[{\citenamefont{Berger}(2003)}]{panoramic}
\bibinfo{author}{\bibfnamefont{M.}~\bibnamefont{Berger}},
  \emph{\bibinfo{title}{A Panoramic View of Riemannian Geometry}}
  (\bibinfo{publisher}{Springer, Berlin}, \bibinfo{year}{2003}).

\bibitem[{\citenamefont{Goldman}(2005)}]{gol05}
\bibinfo{author}{\bibfnamefont{R.}~\bibnamefont{Goldman}},
  \bibinfo{journal}{Comput. Aided Geom. Des.} \textbf{\bibinfo{volume}{22}},
  \bibinfo{pages}{632} (\bibinfo{year}{2005}).

\bibitem[{\citenamefont{Giga}(2006)}]{surface}
\bibinfo{author}{\bibfnamefont{Y.}~\bibnamefont{Giga}},
  \emph{\bibinfo{title}{Surface Evolution Equations: A Level Set Approach}}
  (\bibinfo{publisher}{Birkh{\"a}user, Basel}, \bibinfo{year}{2006}).

\bibitem[{\citenamefont{Landau and Lifshitz}(1958)}]{LLstat}
\bibinfo{author}{\bibfnamefont{L.~D.} \bibnamefont{Landau}} \bibnamefont{and}
  \bibinfo{author}{\bibfnamefont{E.~M.} \bibnamefont{Lifshitz}},
  \emph{\bibinfo{title}{Statistical Physics {\rm [Eq. (33.14)]}}}
  (\bibinfo{publisher}{Pergamon, Oxford}, \bibinfo{year}{1958}).

\bibitem[{\citenamefont{Rugh}(1997)}]{rug97}
\bibinfo{author}{\bibfnamefont{H.~H.} \bibnamefont{Rugh}},
  \bibinfo{journal}{Phys. Rev. Lett.} \textbf{\bibinfo{volume}{78}},
  \bibinfo{pages}{772} (\bibinfo{year}{1997}).

\bibitem[{\citenamefont{Powles et~al.}(2005)\citenamefont{Powles, Rickayzen,
  and Heyes}}]{pow05}
\bibinfo{author}{\bibfnamefont{J.~G.} \bibnamefont{Powles}},
  \bibinfo{author}{\bibfnamefont{G.}~\bibnamefont{Rickayzen}},
  \bibnamefont{and} \bibinfo{author}{\bibfnamefont{D.~M.} \bibnamefont{Heyes}},
  \bibinfo{journal}{Mol. Phys.} \textbf{\bibinfo{volume}{103}},
  \bibinfo{pages}{1361} (\bibinfo{year}{2005}).

\bibitem[{\citenamefont{Schweizer}(2007)}]{sch07}
\bibinfo{author}{\bibfnamefont{K.~S.} \bibnamefont{Schweizer}},
  \bibinfo{journal}{J. Chem. Phys.} \textbf{\bibinfo{volume}{127}},
  \bibinfo{pages}{164506} (\bibinfo{year}{2007}).

\bibitem[{\citenamefont{Tripathy and Schweizer}(2009)}]{tri09}
\bibinfo{author}{\bibfnamefont{M.}~\bibnamefont{Tripathy}} \bibnamefont{and}
  \bibinfo{author}{\bibfnamefont{K.~S.} \bibnamefont{Schweizer}},
  \bibinfo{journal}{J. Chem. Phys.} \textbf{\bibinfo{volume}{130}},
  \bibinfo{pages}{244907} (\bibinfo{year}{2009}).

\bibitem[{\citenamefont{Tripathy and Schweizer}(2011)}]{tri11}
\bibinfo{author}{\bibfnamefont{M.}~\bibnamefont{Tripathy}} \bibnamefont{and}
  \bibinfo{author}{\bibfnamefont{K.~S.} \bibnamefont{Schweizer}},
  \bibinfo{journal}{Phys. Rev. E} \textbf{\bibinfo{volume}{83}},
  \bibinfo{pages}{041407} (\bibinfo{year}{2011}).

\bibitem[{\citenamefont{Lipschitz}(1873)}]{lip1873}
\bibinfo{author}{\bibfnamefont{R.}~\bibnamefont{Lipschitz}},
  \bibinfo{journal}{Bull. Sci. Math. et Astron.} \textbf{\bibinfo{volume}{4}},
  \bibinfo{pages}{297} (\bibinfo{year}{1873}).

\bibitem[{\citenamefont{L{\"u}tzen}(1995)}]{lut95}
\bibinfo{author}{\bibfnamefont{J.}~\bibnamefont{L{\"u}tzen}},
  \bibinfo{journal}{Arch. Hist. Exact Sci.} \textbf{\bibinfo{volume}{49}},
  \bibinfo{pages}{1} (\bibinfo{year}{1995}).

\bibitem[{\citenamefont{Flynn}(1968)}]{fly68}
\bibinfo{author}{\bibfnamefont{C.~P.} \bibnamefont{Flynn}},
  \bibinfo{journal}{Phys. Rev.} \textbf{\bibinfo{volume}{171}},
  \bibinfo{pages}{682} (\bibinfo{year}{1968}).

\bibitem[{\citenamefont{Hall and Wolynes}(1987)}]{hal87}
\bibinfo{author}{\bibfnamefont{R.~W.} \bibnamefont{Hall}} \bibnamefont{and}
  \bibinfo{author}{\bibfnamefont{P.~G.} \bibnamefont{Wolynes}},
  \bibinfo{journal}{J. Chem. Phys.} \textbf{\bibinfo{volume}{86}},
  \bibinfo{pages}{2943} (\bibinfo{year}{1987}).

\bibitem[{\citenamefont{K{\"o}hler and Herzig}(1988)}]{koh88}
\bibinfo{author}{\bibfnamefont{U.}~\bibnamefont{K{\"o}hler}} \bibnamefont{and}
  \bibinfo{author}{\bibfnamefont{C.}~\bibnamefont{Herzig}},
  \bibinfo{journal}{Philos. Mag. A} \textbf{\bibinfo{volume}{58}},
  \bibinfo{pages}{769} (\bibinfo{year}{1988}).

\bibitem[{\citenamefont{Buchenau and Zorn}(1992)}]{buc92}
\bibinfo{author}{\bibfnamefont{U.}~\bibnamefont{Buchenau}} \bibnamefont{and}
  \bibinfo{author}{\bibfnamefont{R.}~\bibnamefont{Zorn}},
  \bibinfo{journal}{Europhys. Lett.} \textbf{\bibinfo{volume}{18}},
  \bibinfo{pages}{523} (\bibinfo{year}{1992}).

\bibitem[{\citenamefont{Heuer and Spiess}(1994)}]{heu94}
\bibinfo{author}{\bibfnamefont{A.}~\bibnamefont{Heuer}} \bibnamefont{and}
  \bibinfo{author}{\bibfnamefont{H.~W.} \bibnamefont{Spiess}},
  \bibinfo{journal}{J. Non-Cryst. Solids} \textbf{\bibinfo{volume}{176}},
  \bibinfo{pages}{294} (\bibinfo{year}{1994}).

\bibitem[{\citenamefont{Sokolov et~al.}(1994)\citenamefont{Sokolov, Kisliuk,
  Quitmann, Kudlik, and R{\"o}ssler}}]{sok94}
\bibinfo{author}{\bibfnamefont{A.~P.} \bibnamefont{Sokolov}},
  \bibinfo{author}{\bibfnamefont{A.}~\bibnamefont{Kisliuk}},
  \bibinfo{author}{\bibfnamefont{D.}~\bibnamefont{Quitmann}},
  \bibinfo{author}{\bibfnamefont{A.}~\bibnamefont{Kudlik}}, \bibnamefont{and}
  \bibinfo{author}{\bibfnamefont{E.}~\bibnamefont{R{\"o}ssler}},
  \bibinfo{journal}{J. Non-Cryst. Solids} \textbf{\bibinfo{volume}{172}},
  \bibinfo{pages}{138} (\bibinfo{year}{1994}).

\bibitem[{\citenamefont{Sanditov and Sangadiev}(1998)}]{san98}
\bibinfo{author}{\bibfnamefont{D.~S.} \bibnamefont{Sanditov}} \bibnamefont{and}
  \bibinfo{author}{\bibfnamefont{S.~S.} \bibnamefont{Sangadiev}},
  \bibinfo{journal}{Glass Phys. Chem.} \textbf{\bibinfo{volume}{24}},
  \bibinfo{pages}{285} (\bibinfo{year}{1998}).

\bibitem[{\citenamefont{Starr et~al.}(2002)\citenamefont{Starr, Sastry,
  Douglas, and Glotzer}}]{sta02}
\bibinfo{author}{\bibfnamefont{F.~W.} \bibnamefont{Starr}},
  \bibinfo{author}{\bibfnamefont{S.}~\bibnamefont{Sastry}},
  \bibinfo{author}{\bibfnamefont{J.~F.} \bibnamefont{Douglas}},
  \bibnamefont{and} \bibinfo{author}{\bibfnamefont{S.~C.}
  \bibnamefont{Glotzer}}, \bibinfo{journal}{Phys. Rev. Lett.}
  \textbf{\bibinfo{volume}{89}}, \bibinfo{pages}{125501}
  (\bibinfo{year}{2002}).

\bibitem[{\citenamefont{Dyre et~al.}(1996)\citenamefont{Dyre, Olsen, and
  Christensen}}]{dyr96}
\bibinfo{author}{\bibfnamefont{J.~C.} \bibnamefont{Dyre}},
  \bibinfo{author}{\bibfnamefont{N.~B.} \bibnamefont{Olsen}}, \bibnamefont{and}
  \bibinfo{author}{\bibfnamefont{T.}~\bibnamefont{Christensen}},
  \bibinfo{journal}{Phys. Rev. B} \textbf{\bibinfo{volume}{53}},
  \bibinfo{pages}{2171} (\bibinfo{year}{1996}).

\bibitem[{\citenamefont{Larini et~al.}(2008)\citenamefont{Larini, Ottochian,
  {De Michele}, and Leporini}}]{lar08}
\bibinfo{author}{\bibfnamefont{L.}~\bibnamefont{Larini}},
  \bibinfo{author}{\bibfnamefont{A.}~\bibnamefont{Ottochian}},
  \bibinfo{author}{\bibfnamefont{C.}~\bibnamefont{{De Michele}}},
  \bibnamefont{and} \bibinfo{author}{\bibfnamefont{D.}~\bibnamefont{Leporini}},
  \bibinfo{journal}{Nature Phys.} \textbf{\bibinfo{volume}{4}},
  \bibinfo{pages}{42} (\bibinfo{year}{2008}).

\bibitem[{\citenamefont{Dyre and Wang}(2012)}]{dyr12}
\bibinfo{author}{\bibfnamefont{J.~C.} \bibnamefont{Dyre}} \bibnamefont{and}
  \bibinfo{author}{\bibfnamefont{W.~H.} \bibnamefont{Wang}},
  \bibinfo{journal}{J. Chem. Phys.} \textbf{\bibinfo{volume}{136}},
  \bibinfo{pages}{224108} (\bibinfo{year}{2012}).

\bibitem[{\citenamefont{Lubchenko and Wolynes}({2007})}]{lub07}
\bibinfo{author}{\bibfnamefont{V.}~\bibnamefont{Lubchenko}} \bibnamefont{and}
  \bibinfo{author}{\bibfnamefont{P.~G.} \bibnamefont{Wolynes}},
  \bibinfo{journal}{Annu. Rev. Phys. Chem.} \textbf{\bibinfo{volume}{{58}}},
  \bibinfo{pages}{235} (\bibinfo{year}{{2007}}).

\bibitem[{\citenamefont{Rabochiy and Lubchenko}(2012)}]{rab12}
\bibinfo{author}{\bibfnamefont{P.}~\bibnamefont{Rabochiy}} \bibnamefont{and}
  \bibinfo{author}{\bibfnamefont{V.}~\bibnamefont{Lubchenko}},
  \bibinfo{journal}{J. Phys. Chem. B} \textbf{\bibinfo{volume}{116}},
  \bibinfo{pages}{5729} (\bibinfo{year}{2012}).

\bibitem[{\citenamefont{Veldhorst et~al.}(2012)\citenamefont{Veldhorst,
  B{\o}hling, Dyre, and Schr{\o}der}}]{vel12}
\bibinfo{author}{\bibfnamefont{A.~A.} \bibnamefont{Veldhorst}},
  \bibinfo{author}{\bibfnamefont{L.}~\bibnamefont{B{\o}hling}},
  \bibinfo{author}{\bibfnamefont{J.~C.} \bibnamefont{Dyre}}, \bibnamefont{and}
  \bibinfo{author}{\bibfnamefont{T.~B.} \bibnamefont{Schr{\o}der}},
  \bibinfo{journal}{Eur. Phys. J. B} \textbf{\bibinfo{volume}{85}},
  \bibinfo{pages}{21} (\bibinfo{year}{2012}).

\bibitem[{\citenamefont{Smith}(1997)}]{smi97}
\bibinfo{editor}{\bibfnamefont{R.}~\bibnamefont{Smith}}, ed.,
  \emph{\bibinfo{title}{Atomic and Ion Collisions in Solids and at Surfaces:
  Theory, Simulation and Applications}} (\bibinfo{publisher}{Cambridge
  University Press}, \bibinfo{year}{1997}).

\end{thebibliography}

\end{document}